\documentclass[preprint]{aastex}
\usepackage{natbib}
\usepackage{emulateapj5}
\shorttitle{Glitches in the Crab Pulsar}
\shortauthors{Wong, Backer, \& Lyne}

\slugcomment{To appear in the Astrophysical Journal}
 
\begin{document}
 
\title{Observations of a Series of Six Recent Glitches in the Crab Pulsar}
\author{Tony Wong and D. C. Backer}
\affil{Astronomy Department and Radio Astronomy Laboratory, University of
California, Berkeley, CA 94720}
\email{twong@astro.berkeley.edu, dbacker@astro.berkeley.edu}
\and
\author{A. G. Lyne}
\affil{University of Manchester, Jodrell Bank Observatory, 
Macclesfield, Cheshire, SK11 9DL, United Kingdom} 
 
\begin{abstract}

From 1995 to 1999, daily monitoring of the radio emission from the
Crab pulsar at the Green Bank and Jodrell Bank observatories revealed
a series of six sudden rotational spinups or glitches, doubling the
number of glitches observed for this pulsar since 1969.  With these
observations, the range of time intervals between significant Crab
glitches has widened considerably, indicating that the occurrence of
Crab glitches may be more random than previously thought. The new
glitch amplitudes $\Delta\nu/\nu$ span an order of magnitude from $2
\times 10^{-9}$ to $3 \times 10^{-8}$. Except in one case, which we
suggest may represent an ``aftershock'' event, the frequency jumps
display an exponential recovery with a timescale of $\sim$3 days for
the smaller glitches and $\sim$10 days for the largest (1996)
glitch. In the largest event, a portion of the spinup was resolved in
time, as was previously reported for the 1989 glitch. A pronounced
change in $\dot{\nu}$ also occurs after each glitch and is correlated
with the size of the initial frequency jump, although for some of the
smaller glitches this appears to be a temporary effect. We discuss the
properties of the ensemble of observed Crab glitches and compare them
with the properties of Vela glitches, highlighting those differences
which must be explained by evolutionary models.

\end{abstract}


\section{INTRODUCTION\label{intro}}

Pulsar timing observations reveal that sudden increases in rotation
rate, known as {\it glitches}, are a common feature in some pulsars,
especially younger ones \citep{She96}. These spinup events are
superposed on the long-term rotational slowdown of the pulsar due to
magnetic dipole radiation and particle outflow, and are usually
followed by a relaxation towards the extrapolated pre-glitch frequency
over a period of days to weeks. Continuous monitoring of the Crab
pulsar (PSR B0531+21) uncovered six significant glitches from 1969 to
1992, at intervals of 3--6 years \citep*{LPS93}. High glitch activity
has been observed in other pulsars as well. For example, at least
twelve glitches of the Vela pulsar (PSR B0833$-$45) were observed
between 1969 and 1996 \citep[e.g.,][]{Chau93,Fla95}, usually at
intervals of 2--3 years, and twelve glitches have been detected
in the rotation of PSR J1341$-$6220 during just 8.2 years of
observations \citep{Wang00}, making it the most frequently glitching
pulsar known.

At the same time, precise timing measurements have revealed that many
pulsars exhibit an unpredictable fluctuation in pulse arrival time known as
{\it timing noise}. A number of previous studies \citep{Boy72,Gro75,Cor80}
have shown that timing noise in the Crab pulsar can be modeled as a random
walk in frequency. The individual steps in this random walk have not been
resolved, but can be of either sign and have amplitudes $|\Delta\nu/\nu|
\ll 10^{-10}$, whereas major glitches of the Crab can be distinguished as
sudden positive jumps in frequency with $\Delta\nu/\nu \gtrsim 10^{-9}$,
followed by an exponential recovery over a period of days to months. Still,
the possibility remains that glitches lie at the upper end of a continuous
spectrum of frequency jumps \citep{CH80}, and careful consideration of
observational selection effects is essential before attempting to draw
general conclusions about glitches.

Although our understanding of the glitch phenomenon is limited by our
incomplete knowledge of neutron star structure, the most plausible
explanation for the frequency spinups is the dumping of
angular momentum from an unseen reservoir into the part of the star whose
rotation we measure \citep[e.g.,][]{Alp93}. Since the outer crust and core
superfluid are believed to be coupled on short ($<$1 minute) timescales to
the magnetic field and the observed pulsar beam (e.g., \citealt*{Alp84};
\citealt{Mend98}), the probable site of this reservoir is a superfluid
component in the inner crust. As discussed by \citet{And75} and others, the
quantized vortex lines which carry angular momentum in the superfluid can
be ``pinned'' to lattice sites in the neutron-rich inner crust, inhibiting
the outward motion of vortex lines that occurs as the star spins down and
creating an excess rotation of the superfluid relative to the outer crust.
If large numbers of pinned vortices were to simultaneously unpin, the
excess angular momentum would be shared with the rest of the star and a
glitch would be observed. While this scenario provides an attractive
theoretical framework for understanding glitches, the mechanisms that 
trigger glitches and lead to the observed postglitch recovery remain
controversial \citep[e.g.,][]{Alp96,Lin96,Jon98}.

Observations of pulsar glitches, in addition to providing insights into
the glitch phenomenon itself, offer one of the few direct probes of
neutron star structure and the physics of ultradense matter. For
instance, the frequent large glitches ($\Delta\nu/\nu \sim 10^{-6}$)
seen in the Vela pulsar indicate that the superfluid component in the
inner crust comprises $\sim$2\% of the star's moment of inertia (e.g.,
\citealt{Chau93}; \citealt*{Lin99}). Crab glitches, on the other hand,
exhibit much smaller relative frequency jumps ($\Delta\nu/\nu \sim
10^{-8}$), and tend to be dominated by persistent changes in the
slowdown rate with $\Delta\dot{\nu}/\dot{\nu} \sim 10^{-4}$
\citep[e.g.,][]{Gull77,Dem83}.  The persistent shift in $\dot{\nu}$ has
been attributed to either a change in the external torque
\citep*{Lin92} or a change in the moment of inertia acted on by the
torque \citep{Alp96}. The various exponential relaxation timescales
that follow glitches have been interpreted, in the framework of the
``vortex creep'' theory, as resulting from distinct regimes of
superfluid pinning to the crustal lattice \citep{Alp93}.

Since the occurrence of glitches is unpredictable and their relaxation
timescales can be quite short (a few days or less), daily monitoring of
frequent glitchers such as the Crab and Vela pulsars are necessary in
order to study the glitch process in detail.  Here we present radio
timing observations taken during a period of increased glitching in the
Crab that began in 1995 and has extended into 2000.  The data were
collected as part of regular pulsar monitoring programs with dedicated
small telescopes at the NRAO\footnote{The National Radio Astronomy
Observatory (NRAO) is operated by Associated Universities, Inc, with
funding from the National Science Foundation.} Green Bank and the
University of Manchester Jodrell Bank sites.  The data acquisition and
reduction procedures are described in \S 2, and in \S 3 we present our
analysis of six distinct frequency jumps during this period.  A large
glitch was observed in 1996 June and smaller events were observed in
1995 October, 1997 January, February, and December, and 1999 October.
(The recent large glitch on 2000 July 15 is not covered by these
observations).  In \S 4 we examine the frequency (rate of occurrence)
of Crab glitches, and \S 5 we discuss the observed properties of the
glitches.  Our conclusions are summarized in \S 6.  Throughout this
paper we define the rotation frequency $\nu$ in Hz, which can be
converted to angular frequency using $\Omega$=2$\pi\nu$.

%%%%%%%%%%%%%%%%%%%%%%%%%%%%%%%%%%%%%%%%%%%%%%%%%%%%%%%%%%%%%%%%%%%%%%%%%%%%%
\bigskip

\section{OBSERVATIONS AND DATA REDUCTION}\label{observ}

Observations at the Green Bank (hereafter GB) site of the NRAO are
conducted as part of a regular monitoring program on a 26 m (85 ft)
radio telescope.  The Crab pulsar is observed daily, with typically
eight 10-minute integrations at 610 MHz and three integrations at 327
MHz.  Two linear polarizations are measured separately and combined
later in the data analysis software.  A convolution processor
dedisperses the signal in real time within each of 32 frequency
channels of 0.5 (0.25) MHz bandwidth at 610 (327) MHz, thus correcting
for pulse smearing across each channel due to interstellar dispersion.
The signal is then averaged synchronously with the predicted apparent
pulse period into approximately 1000 time bins (1 bin $\approx$ 34
$\mu$s).  A time stamp accurate to $\sim0.1$ $\mu$s is recorded at the
start of each integration.  For calibration purposes, integrations on 
a standard noise source are performed before the source integrations.

In software, the integrated profiles from all 32 channels are aligned
using the assumed dispersion measure (DM), flux calibrated, and summed
to form a single pulse profile for each integration.  
A high signal-to-noise average profile from several integrations is
used as a template, and a simple cross-correlation is performed to
determine the arrival time (TOA) of the main pulse peak.  A mean TOA
for each day or half-day is then determined using a least-squares
linear fit to all of the TOA's during that period, as in
\citet{She96}.  An associated uncertainty is also derived
from this least-squares fit.  The typical uncertainty in the
day-averaged TOA at 610 MHz is 10 $\mu$s, but increases to around 60
$\mu$s during periods of heightened interstellar scattering, as
indicated by broadening of the 327 MHz pulse profiles (see below).
Arrival times are then corrected to the solar system barycenter using
the standard TEMPO software package \citep{Tay89}.

Observations at the University of Manchester's Jodrell Bank Observatory
(hereafter JB) are conducted at 610 MHz with a 12.5-m telescope as part
of a long-term program of monitoring the Crab pulsar. The pulsar is
observed daily over a timespan of up to 14 hours; a full description of
the data acquisition and reduction techniques can be found in
\citet{LPS93}.  The 610-MHz data are supplemented by occasional
observations at 1400 MHz using the 76-m Lovell telescope, in order to
monitor changes in DM.  Because the on-source integration times with
the 12.5-m telescope are much longer than for the GB observations, the
uncertainties in the day-averaged TOA's are comparable to those for the
GB data ($\sim$ 10 $\mu$s).  There are, however, variations of up to
$\sim$50 $\mu$s in the offset between the two datasets during periods
of increased scattering, probably a result of changes in the observed
pulse profile.  Thus when combining the datasets, we restricted
ourselves to periods of minimal scattering during which the offset was
roughly constant.

%%%%%%%%%%%%%%%%%%%%%%%%%%%%%%%%%%%%%%%%%%%%%%%%%%%%%%%%%%%%%
%%%%%%%%%%%%%%%%%%%%%%%%   FIG. 1   %%%%%%%%%%%%%%%%%%%%%%%%%
%%%%%%%%%%%%%%%%%%%%%%%%%%%%%%%%%%%%%%%%%%%%%%%%%%%%%%%%%%%%%

\vskip 0.4truein
\includegraphics[width=3.3in]{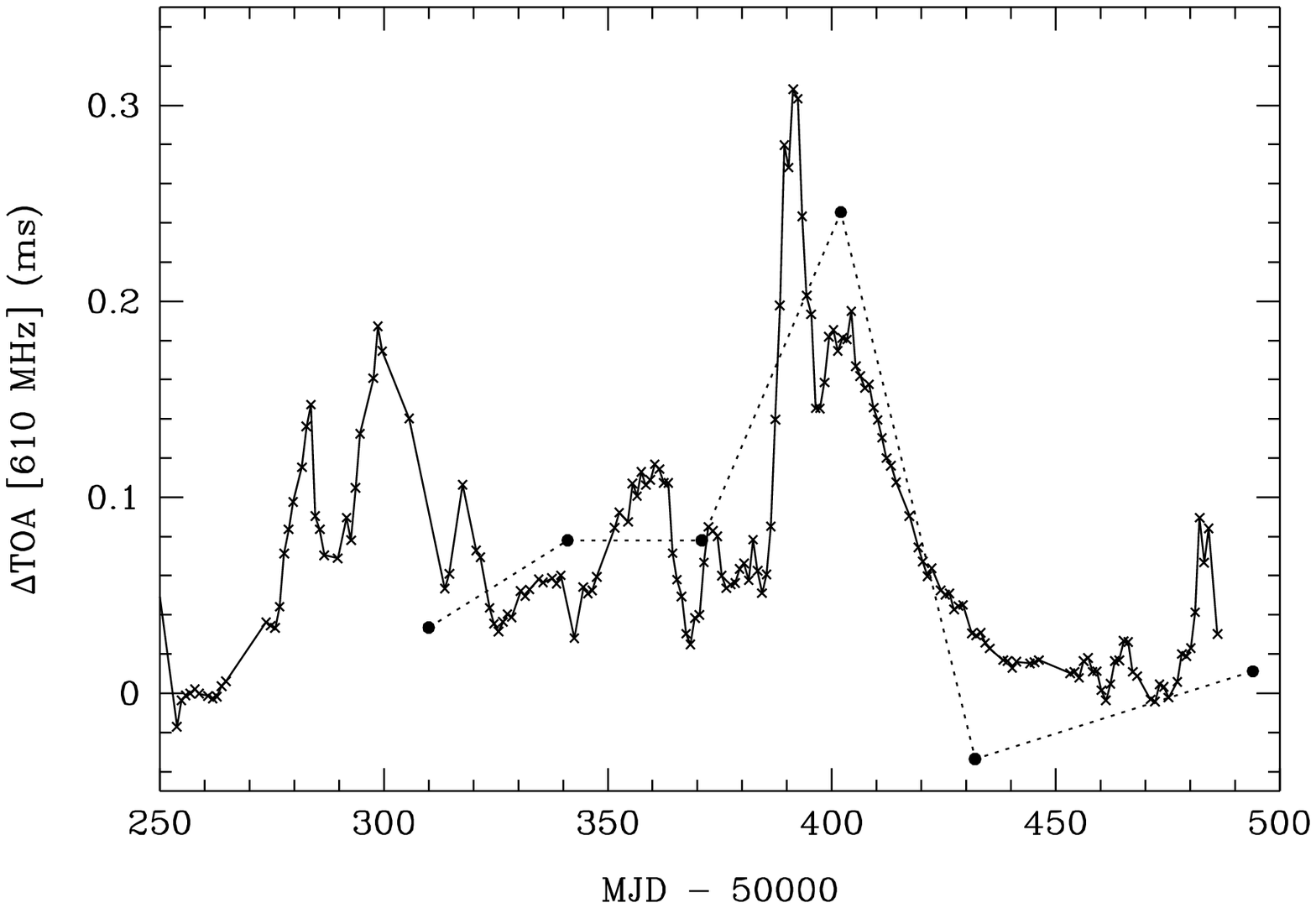}
\figcaption[dmerr.eps]{
Estimated TOA fluctuations at 610 MHz due to changes in dispersion
measure (DM).  The DM fluctuations are determined by comparison of the
610 MHz and 327 MHz pulse arrival times, after the 327 MHz data have
been corrected for scattering (see text).  Since we have not determined
absolute arrival times, the zero point of the y-axis is defined at MJD
50259.  The filled circles are points from the Jodrell Bank DM
monitoring programme, with a reference DM of 56.82 chosen to produce a
rough match with the GB data.
\label{dmerrfig}}
\vskip 0.25truein

In this paper we confine our analysis to the 610 MHz data, principally from
Green Bank but with the JB data included during periods of sparse sampling.
The 327 MHz GB data proved to be of lower quality for timing purposes, but
were useful for monitoring changes in scattering and dispersion
\citep*[e.g.,][]{Bac00}, which can produce systematic errors in the 610-MHz
TOA's. Interstellar scattering leads to pulse broadening due to multipath
propagation, which can be modeled by assuming that the intrinsic pulse
shapes are Gaussian and convolving the profile with a one-sided exponential
decay \citep*[e.g.,][]{Kom72}. Variable scattering was found to be
important at 327 MHz but generally not at 610 MHz, due to its strong
frequency dependence.  Changes in DM were monitored by correcting the 327
MHz data for scattering using the exponential model, then taking the
difference of the 610 MHz and 327 MHz timing residuals with a common
spindown model applied to both data sets. From these measurements we
inferred that DM changes can contribute offsets of up to $\sim$100 $\mu$s
to the 610 MHz TOA's, although the changes generally occur gradually over
several weeks (Figure~\ref{dmerrfig}). As an example, a DM change of 10
$\mu$s per day leads to an error in the rotation frequency of $\sim 5
\times 10^{-9}$ Hz, which is at least an order of magnitude smaller than
the glitch events described here. In light of the uncertainties in the
scattering correction and a lack of daily DM measurements during some
epochs, we have generally chosen not to apply a DM correction to our timing
data. However, in the months following 1997 October ($\sim$ MJD 50723) the
DM was found to change extremely rapidly (\S\ref{dmjump}), and for Glitch
11 (\S\ref{g11}) we have attempted to include these effects in our
modeling.

%%%%%%%%%%%%%%%%%%%%%%%%%%%%%%%%%%%%%%%%%%%%%%%%%%%%%%%%%%%%%%%%%%%%%%%%%%%%%

\section{RESULTS OF GLITCH MODEL FITTING}\label{results}

\begin{figure*}
\begin{center}
\includegraphics[width=7in]{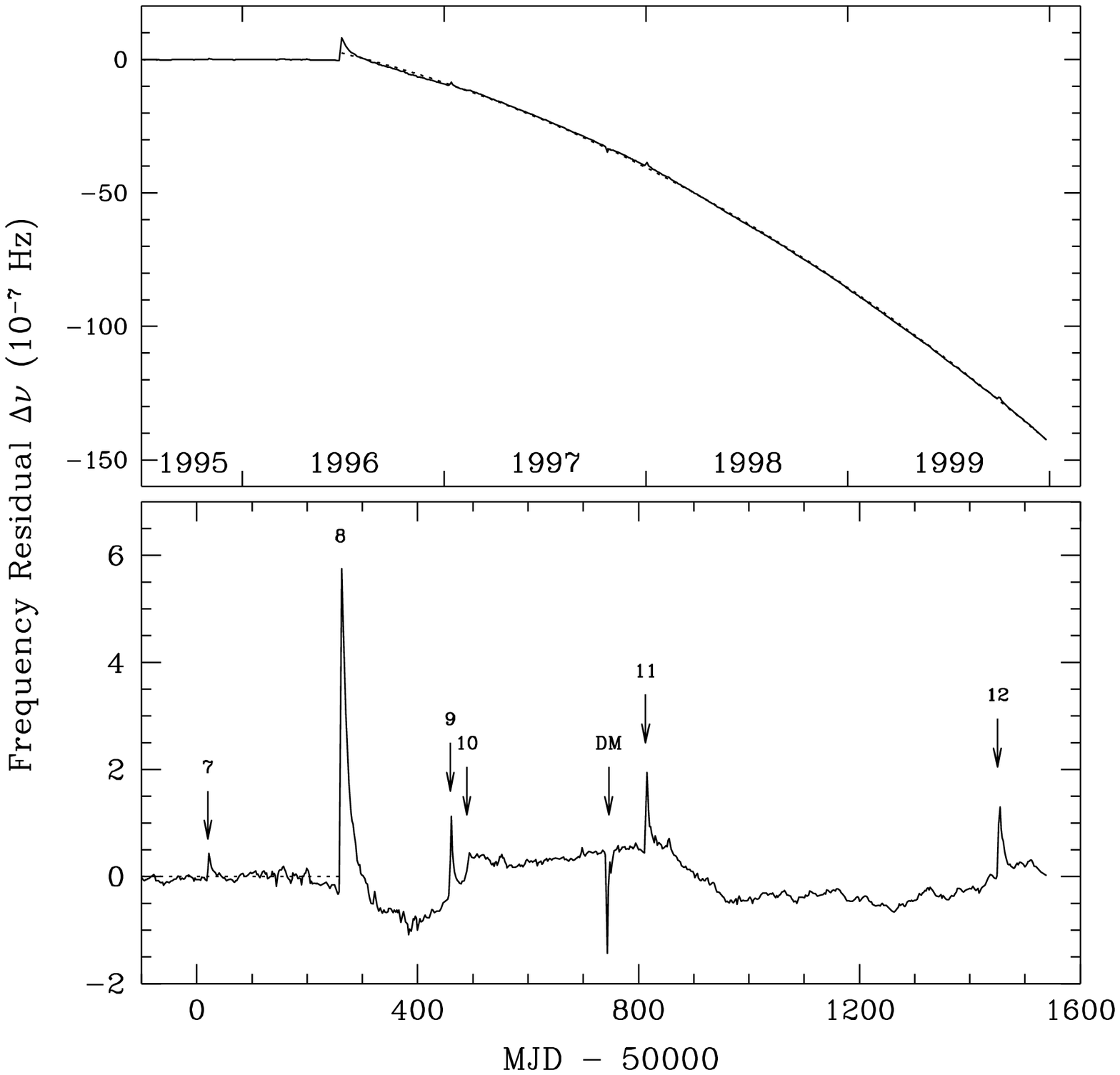}
\vskip 0.25truein
\figcaption[nurecord.eps]{
Frequency residuals derived from GB data, after subtraction of a model
fit to the 500 days before the 1996 June glitch.  In the bottom panel, a
quadratic fit to the residuals from MJD 50350--51400 has been removed
(dotted line in the top panel) in order to highlight discrete events.
\label{nurecord}}
\end{center}
\end{figure*}

For the study of both glitches and timing noise, one commonly fits to
the arrival-time data (expressible as rotational phase $\phi$) a
simple spindown model of the form:
\begin{equation}
\phi_m(t)=\phi_0+\nu_0 (t-t_0) + \frac{1}{2}\dot{\nu}_0 (t-t_0)^2 +
\frac{1}{6}\ddot{\nu}_0 (t-t_0)^3\;,
\end{equation}
where $\phi_m(t)$ is the predicted pulse phase at time $t$ and $t_0$
is an arbitrary reference time. Timing irregularities appear as {\it
phase residuals} ($\phi-\phi_m$), deviations of the observed pulse
phase from the predicted phase. {\it Frequency residuals} ($\nu-\nu_m
= \dot{\phi} - \dot{\phi}_m$) are then calculated by averaging the
phase residuals over $\sim 1$ day intervals and determining the slope
between adjacent points.  We estimate that the typical error in the
frequency determinations is $5 \times 10^{-9}$ Hz except during
periods of severe scattering.  In Figure~\ref{nurecord} (upper panel)
the frequency residuals from the GB data are shown, based on a model
fitted to the JB data over the 500 days prior to the 1996 glitch.  The
values of $\nu$, $\dot{\nu}$, and $\ddot{\nu}$ just before this
glitch, as determined by the model, are given in Table~\ref{paramtbl}.
In the lower panel of Fig.~\ref{nurecord} a quadratic fit to the
residuals following this glitch (corresponding to jumps in $\nu$,
$\dot{\nu}$, and $\ddot{\nu}$) has been removed to permit closer
inspection of the residuals.

After subtracting a preglitch spindown model given by Equation (1), we
characterize a sudden glitch event by permanent jumps in $\nu$,
$\dot{\nu}$, and $\ddot{\nu}$ accompanied by exponentially decaying
jumps in $\nu$.  \citet{She96} have applied such a model to a large
sample of glitching pulsars.  If $t$ is the time since the glitch,
then the postglitch frequency residuals can be written as:
\begin{equation}
\Delta\nu = \Delta\nu_p + \Delta\dot{\nu}_p t + 
	\frac{1}{2}\Delta\ddot{\nu}_p t^2 + 
	\sum_n \Delta\nu_n\,e^{-t/\tau_n}\;.
\end{equation}
Thus, at the glitch time ($t=0$) the initial unresolved step in frequency
is given by
\begin{equation}
\Delta\nu_0 = \Delta\nu_p + \sum_n \Delta\nu_n \,\,{\rm .}
\end{equation}
To minimize the number of fitting variables, we have assumed that all 
postglitch components begin simultaneously (at $t=0$).

%%%%%%%%%%%%%%%%%%%%%%%%%%%%%%%%%%%%%%%%%%%%%%%%%%%%%%%%%%%%%
%%%%%%%%%%%%%%%%%%%%%%%%   FIG. 2   %%%%%%%%%%%%%%%%%%%%%%%%%
%%%%%%%%%%%%%%%%%%%%%%%%%%%%%%%%%%%%%%%%%%%%%%%%%%%%%%%%%%%%%

In order to take advantage of all available data, the glitch model was
rewritten in terms of rotational phase and fitted directly to the
TOA's that had been averaged into 0.5--1 d intervals.  Results of the
fitting are summarized in Table~\ref{fittbl}.  For each glitch up to
three models are shown, each employing a different parameter set or
range of data; the maximum length of the data span is prescribed by
the interval between glitches.  The nonlinear least-squares fitting
routine we used seeks to minimize the reduced $\chi^2$, defined as
\begin{equation}
\tilde\chi^2(\phi_m') = \frac{1}{N-M}\;\sum_{i=1}^N
	\left(\frac{\phi-\phi_m'}{\sigma_i}\right)^2,
\end{equation}
where $N$ is the number of points in the fit, $M$ is the number of fit
parameters, $\phi_m'$ is the model phase (including the glitch model),
and $\sigma_i$ is the uncertainty in each phase measurement (\S 2).
Also shown in Table~\ref{fittbl} is $\tilde\chi^2(\nu_m')$, which
describes how well the model {\it frequencies} fit the observed
frequencies.  Each transient term in the fit has been labeled with
subscript 1, 2, or 3 based on whether it represents a short-term
($\lesssim$50~d) exponential decline ($\Delta\nu_n > 0$), long-term
($\gtrsim$50~d) exponential rise or decline, or very short-term
($\lesssim$1~d) exponential rise ($\Delta\nu_n < 0$), respectively.

Our adopted ``best fits,'' identified with asterisks in
Table~\ref{fittbl}, are presented in Table~\ref{alltbl}, along with
the corresponding parameters for the previously observed Crab
glitches.  (The individual glitches are discussed in more detail
below.)  Parameters of the first five glitches are taken from
\citet{LPS93}, after adjusting their fits to conform with the model
given by Equation (2).  Parameters for Glitch 6 have been derived from
unpublished JB timing data.

The accuracy of the fit parameters is difficult to determine because there
are inevitably correlations between parameters, there is no obvious time
period over which to perform the fit, and the fit is dependent on the
assumed spindown model before the glitch. Our analysis is also susceptible
to systematic errors resulting from DM variations (\S\ref{observ}), and to
intrinsic timing noise in the pulsar which cannot be modeled in simple
terms. The latter will tend to increase the formal $\tilde\chi^2$ as the
timespan of the fit increases. For events prior to 1995, the uncertainties
are those given by \citet{LPS93}. For the 1995--9 events, the quoted
uncertainties are determined by shifting each parameter value above and
below the best-fit value until the resulting value of
$\tilde\chi^2(\nu_m')$ increases by 1. This criterion corresponds to
allowing a $1\sigma$ shift in each frequency measurement away from the
model. In most cases, the other parameters were allowed to vary during this
process. As a consistency check we have fitted the GB and JB data
separately whenever possible; error bars have been extended when necessary
to encompass the resulting fits. The quoted errors do not include the
uncertainty in the preglitch timing model. 

\subsection{Glitch 7: MJD 50020\label{g7}}

This small event ($\Delta\nu/\nu \sim 2 \times 10^{-9}$) is easily
distinguished from the general timing noise when viewed in the frequency
residuals (Fig.~\ref{nurecord}). It may be similar to the small 1971
event reported by \citet{Loh72}, although in that case gaps in the timing
observations make it difficult to establish whether an exponential decay
was present. Due to the small size of the glitch and a systematic variation
by $\sim$10 $\mu$s in the offset between the GB and JB residuals, we have
fitted a glitch model to the JB residuals only rather than attempting to
combine the datasets. Fits to the GB dataset are consistent with the
adopted fit within the stated errors.

%%%%%%%%%%%%%%%%%%%%%%%%%%%%%%%%%%%%%%%%%%%%%%%%%%%%%%%%%%%%%
%%%%%%%%%%%%%%%%%%%%%%%%   FIG. 3   %%%%%%%%%%%%%%%%%%%%%%%%%
%%%%%%%%%%%%%%%%%%%%%%%%%%%%%%%%%%%%%%%%%%%%%%%%%%%%%%%%%%%%%

\vskip 0.25truein
\includegraphics[width=3.3in]{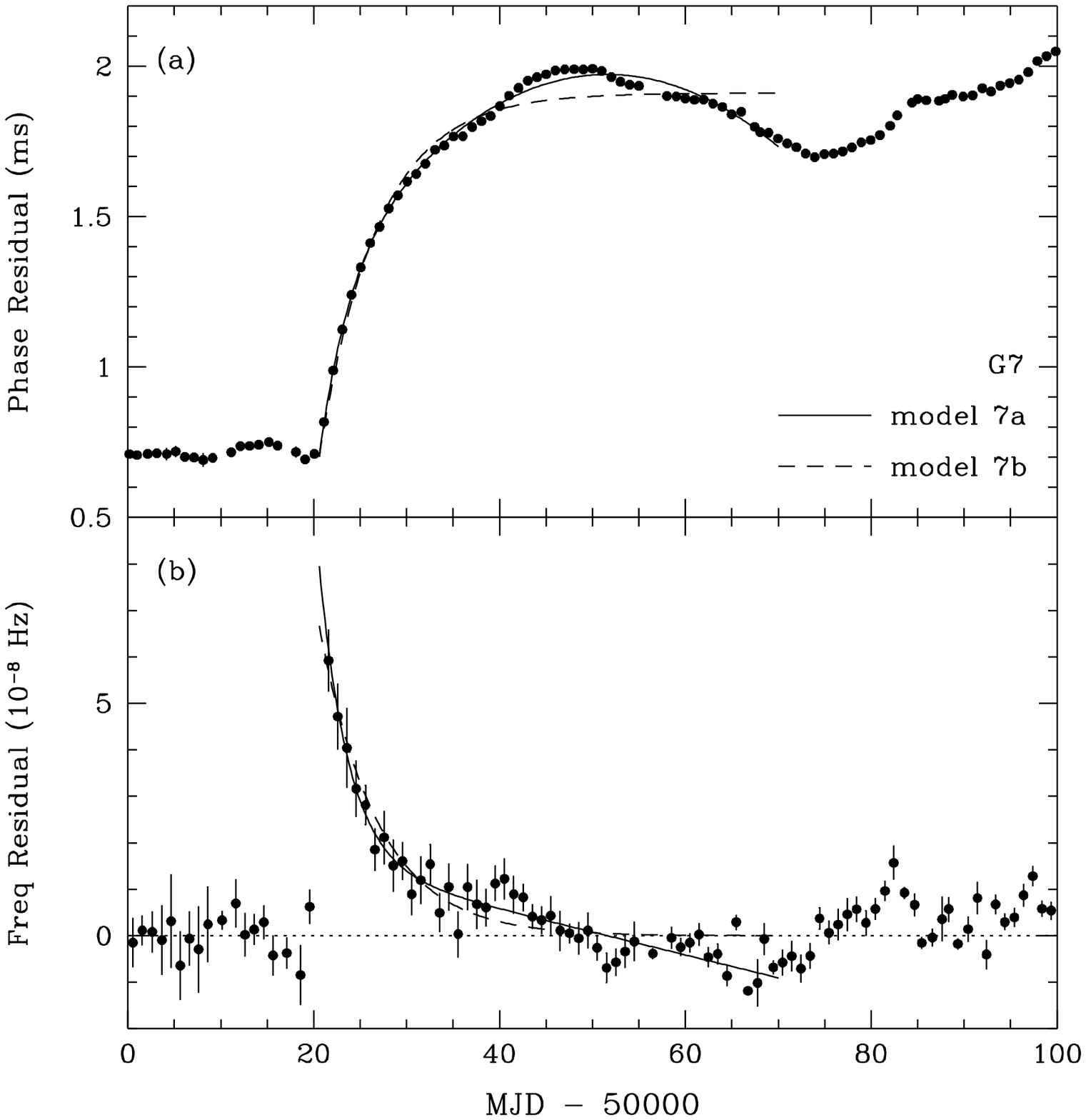}
\figcaption[g7.eps]{
Phase and frequency residuals showing the effects of Glitch 7 in
1995 October.  Two postglitch timing models are shown, one with a
temporary jump in $\dot\nu$ (solid line) and one with only a
transient exponential term (dashed line).
\label{g7fig}}
\vskip 0.25truein

Figure~\ref{g7fig} displays the JB phase residuals in milliseconds
[i.e., $(\phi-\phi_m)P$ with $P=1/\nu$] and the frequency residuals in
units of $10^{-8}$ Hz after removing a spindown model fitted to the
300~d prior to the event.  The solid line is the adopted glitch model
of the form given by Eq.~(2) without the $\Delta\ddot{\nu}$ term,
fitted to the 50 days after the glitch.  Note that the model includes a
temporary change in $\dot{\nu}$ over this timespan that vanishes
around MJD 50075.  The dashed line represents a simpler model
consisting of only a pure exponential decay (no ``persistent'' jump in
$\nu$ or $\dot{\nu}$), which is a much poorer fit to the data more
than 20~d after the glitch.  On the other hand, since the residuals
for even this simple model are not significantly greater than the general
timing noise, such a model could still be a valid description of the
data.

\subsection{Glitch 8: MJD 50260\label{g8}}

On 1996 June 25--26, the pulsar rotation frequency experienced a large
jump of $\Delta\nu/\nu \sim 3 \times 10^{-8}$, of which roughly
$1/3$ may have been resolved in time, as will be argued below. For the
analysis, the preglitch timing model used in Fig.~\ref{nurecord} was
adopted, while a postglitch recovery model was fit to the following
190 days (since the recovery was interrupted by another glitch in
1997 January). The Jodrell Bank data were included only during the periods
MJD 50200--50280, 50325--50375 and 50410--50450, when scattering effects
were judged to have negligible impact on the offset between the two
datasets. The 1$\sigma$ uncertainty in the offset is about 15 $\mu$s.

In addition to permanent jumps in $\nu$ and $\dot{\nu}$ and an
exponential decay with a timescale of $\tau_1 \sim 10$ d, there is a
long-term component that can be modeled either as an exponential decay
with a timescale of $\tau_2 \sim 100$ d, or as a ``permanent'' (up to
the next glitch) positive jump in $\ddot{\nu}$.  As noted by
\citet{She96}, this ambiguity results from the fact that the
exponential time constant $\tau_2$ is comparable to the timespan of
the fit.  While the model with a change in $\ddot{\nu}$ provides a
better fit to the data ($\tilde\chi^2$ of 831 rather than 911), in
both cases $\tilde\chi^2$ is likely dominated by unmodeled timing
noise, so it is unclear whether the difference between the two is
significant.  (For comparison, fitting a spindown model to an
equivalent interval well separated from glitches, from MJD 51000 to
51190, gives $\tilde\chi^2$ of 800).  In Table~\ref{fittbl} we give
the best-fit values corresponding to each of the two models (8a and
8b).  The 100-d exponential decay has a timescale comparable to that
observed in the 1975 and 1989 glitches \citep{LPS93}, but its
amplitude has the opposite sign (i.e.\ it represents a slowdown rather
than a spinup).

%%%%%%%%%%%%%%%%%%%%%%%%%%%%%%%%%%%%%%%%%%%%%%%%%%%%%%%%%%%%%
%%%%%%%%%%%%%%%%%%%%%%%%   FIG. 4   %%%%%%%%%%%%%%%%%%%%%%%%%
%%%%%%%%%%%%%%%%%%%%%%%%%%%%%%%%%%%%%%%%%%%%%%%%%%%%%%%%%%%%%

\vskip 0.25truein
\includegraphics[width=3.3in]{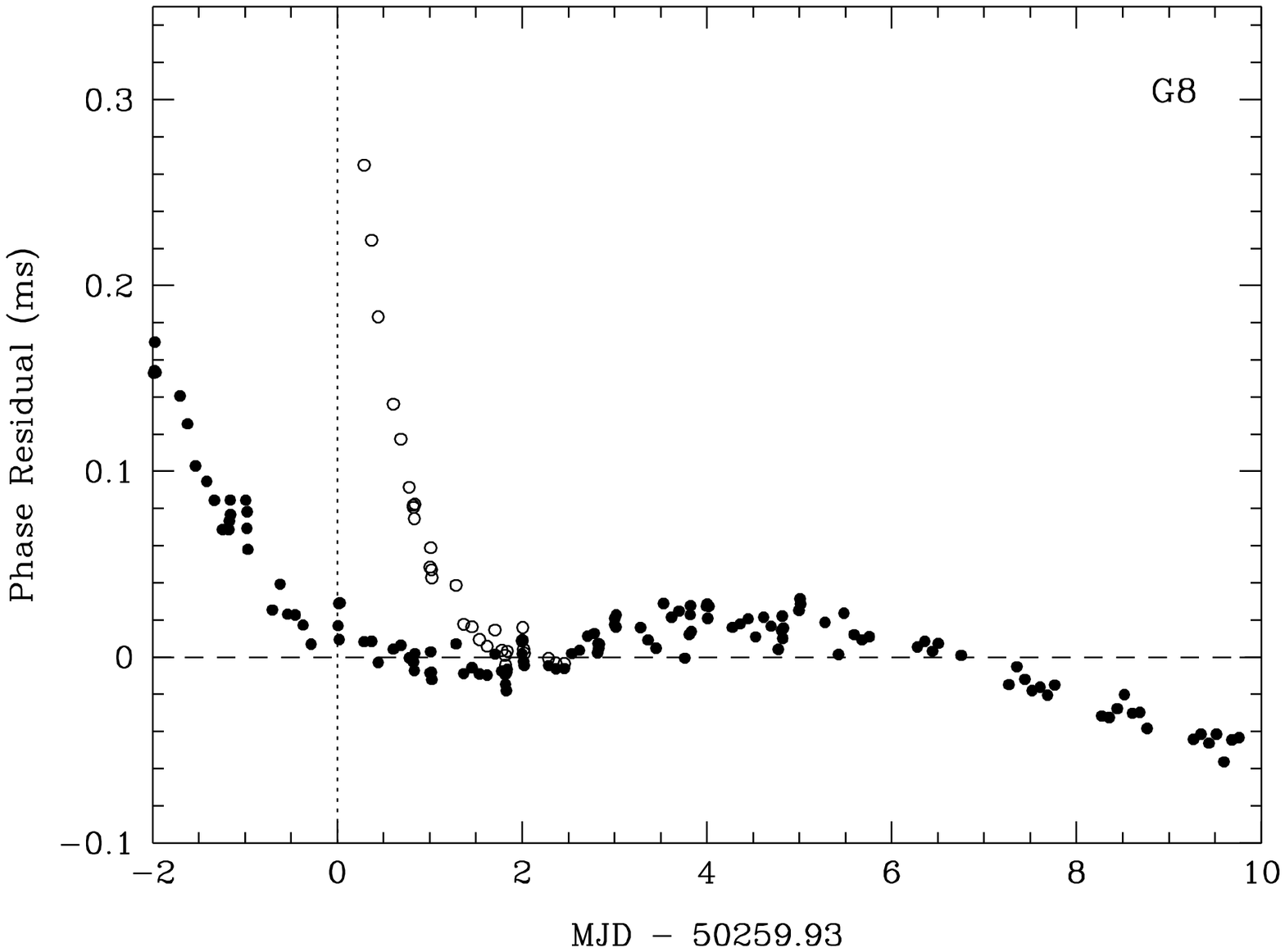}
\figcaption[g8spup.eps]{
Phase residuals immediately after the 1996 glitch (G8), after removal of
persistent changes in $\nu$, $\dot{\nu}$, and $\ddot{\nu}$, a 10.3 d
exponential decay, and a 0.5 d exponential rise.  The open circles are
the phase residuals if the 0.5 d rise is {\it not} included.
The time axis is referenced to the assumed glitch time.
\label{g8spinup}}
\vskip 0.25truein

Upon removing the components discussed above, it becomes clear from the
phase residuals that a short-term spinup of the pulsar occurred in the
first two days (Figure~\ref{g8spinup}).  Although the magnitude of the
residuals is consistent with timing noise alone, their temporal
dependence (with a sudden rise and exponential-like decay) leaves
little doubt that they are associated with the glitch. The fitted
exponential timescale for the spinup is roughly 0.5~d and its amplitude
is about 0.5 $\Delta\nu_1$.  These values are given in
Table~\ref{fittbl} as Model 8c.  In comparison, the spinup time seen in
the 1989 glitch was 0.8~d and its relative amplitude was about 0.3
$\Delta\nu_1$ \citep*{LSP92}. As was the case in 1989, part of the
spinup is still unresolved ($\Delta\nu_0 > 0$).

%%%%%%%%%%%%%%%%%%%%%%%%%%%%%%%%%%%%%%%%%%%%%%%%%%%%%%%%%%%%%
%%%%%%%%%%%%%%%%%%%%%%%%   FIG. 5   %%%%%%%%%%%%%%%%%%%%%%%%%
%%%%%%%%%%%%%%%%%%%%%%%%%%%%%%%%%%%%%%%%%%%%%%%%%%%%%%%%%%%%%

\vskip 0.3truein
\includegraphics[width=3.3in]{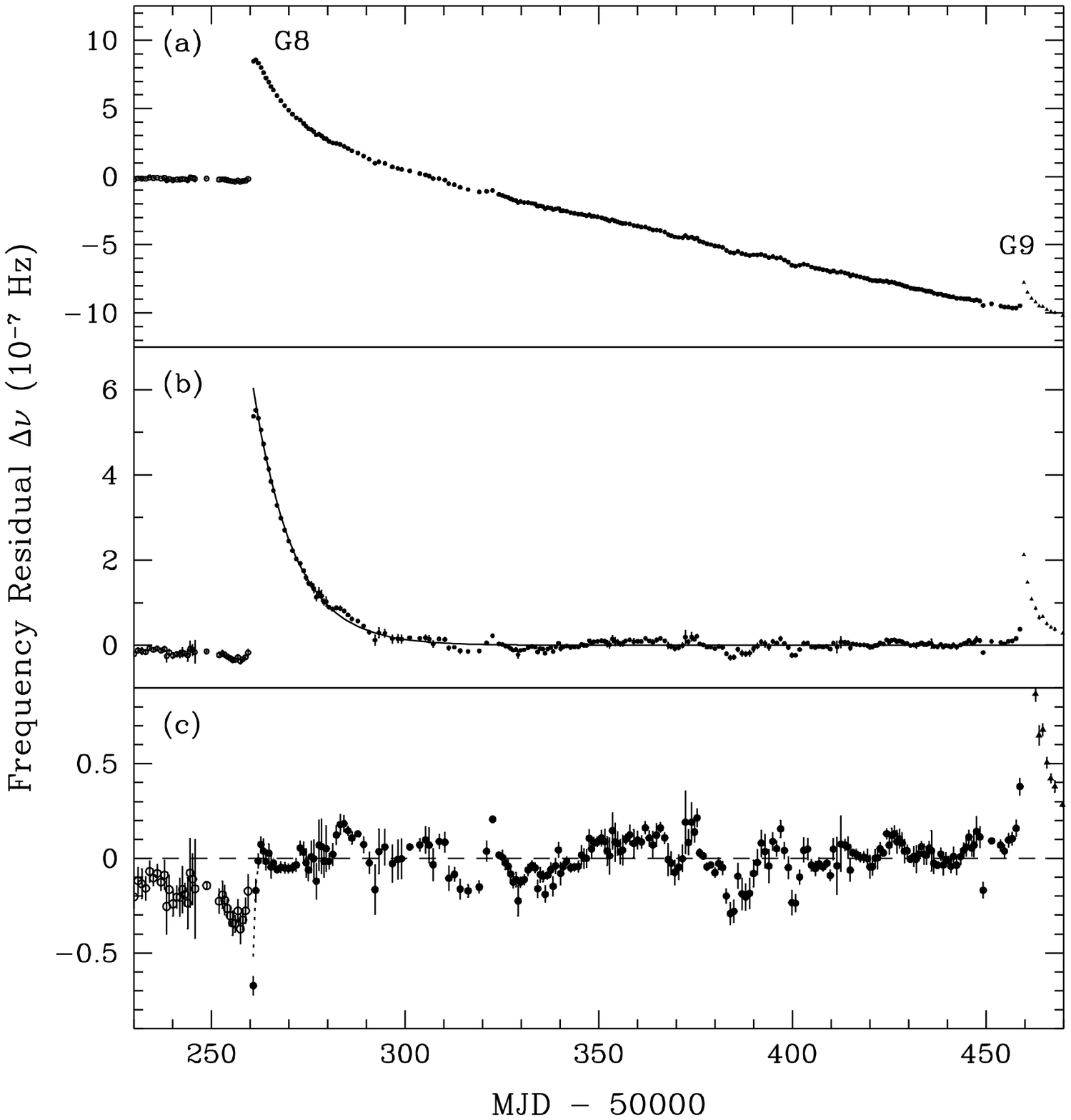}
\figcaption[g8.eps]{
Frequency residuals around the time of the large glitch of 1996 June (G8),
after (a) subtraction of only the preglitch spindown model; (b) removal
of persistent changes in $\nu$, $\dot{\nu}$, and $\ddot{\nu}$ (the
10-d transient is shown as a solid line); (c) removal of the 10-day
transient component as well (the 0.5-d transient is shown as a dotted
line).  Open circles are preglitch points, solid triangles are points after
the 1997 January glitch (G9), and solid circles are points between the two
glitches.
\label{g8fig}}
\vskip 0.3truein

Figure~\ref{g8fig} provides a visual breakdown of the components in
the adopted glitch model, in which we parametrize the long-term
postglitch behavior using a change in $\ddot{\nu}$. We begin with the
frequency residuals after subtraction of the preglitch model
[Fig.~\ref{g8fig}(a)].  Note that after $\sim$50~d the dominant effect
of the glitch is a permanent increase in the spindown rate
$|\dot{\nu}|$, as was seen in the large glitches of 1975 and 1989. In
Fig.~\ref{g8fig}(b), the fitted $\Delta\nu_p$, $\Delta\dot{\nu}_p$ and
$\Delta\ddot{\nu}_p$ after the glitch have been removed. The solid
line represents the fitted $\tau_1 = 10$ d exponential. In
Fig.~\ref{g8fig}(c) the 10~d transient component has also been
removed, revealing the short-term rise following the glitch. Note the
striking slowdown in frequency by $\sim 3 \times 10^{-8}$ Hz that
precedes the glitch, a 5$\sigma$ offset from the preglitch model.
Since this slowdown takes place over a much shorter timescale
($\sim$20~d) than the timescale over which the preglitch model fit can
vary ($\sim$250~d), it appears to be significant.  However, as such a
glitch ``precursor'' is not observed before the subsequent glitch 200
days later, it does not appear to be a general feature of Crab
glitches.

\subsection{Glitches 9 and 10: MJD 50459 and 50489\label{g9_10}}

In early 1997, two glitches occurred roughly 30~d apart
(Figure~\ref{g9fig}).  The first event ($\Delta\nu/\nu \sim 8 \times
10^{-9}$) occurred on MJD 50459 (1997 January 11), and was well
sampled in time, occurring within $\sim$1~hr of observations at GB.
It is dominated by a transient term with a timescale $\tau\approx 3$~d
and shows no evidence for a resolved spinup.  Although there is a
measurable change in $\dot{\nu}$, little can be said about any change
in $\ddot{\nu}$ because the fit was performed over just 30 days.  The
second event ($\Delta\nu/\nu \sim 2 \times 10^{-9}$), which occurred
in early February during a gap in the GB observations, can be modeled
as step changes in $\nu$, $\dot{\nu}$, and $\ddot{\nu}$ based on the
GB data alone (Model 10a in Table~\ref{fittbl}); inclusion of JB data
recorded during the gap indicates a gradual spinup with a timescale of
$\sim$2 d (Model 10b, the adopted fit).  Modeling the long-term
recovery with an asymptotic exponential yields a much poorer fit
(Model 10c).  This model is also poorly constrained because $\tau$ is
much longer than the timespan of the fit, which extends up to the next
glitch (MJD 50812) but excludes a period from MJD 50742-50752 as
described below (\S\ref{dmjump}).

%%%%%%%%%%%%%%%%%%%%%%%%%%%%%%%%%%%%%%%%%%%%%%%%%%%%%%%%%%%%%
%%%%%%%%%%%%%%%%%%%%%%%%   FIG. 6   %%%%%%%%%%%%%%%%%%%%%%%%%
%%%%%%%%%%%%%%%%%%%%%%%%%%%%%%%%%%%%%%%%%%%%%%%%%%%%%%%%%%%%%

\vskip 0.25truein
\includegraphics[width=3.3in]{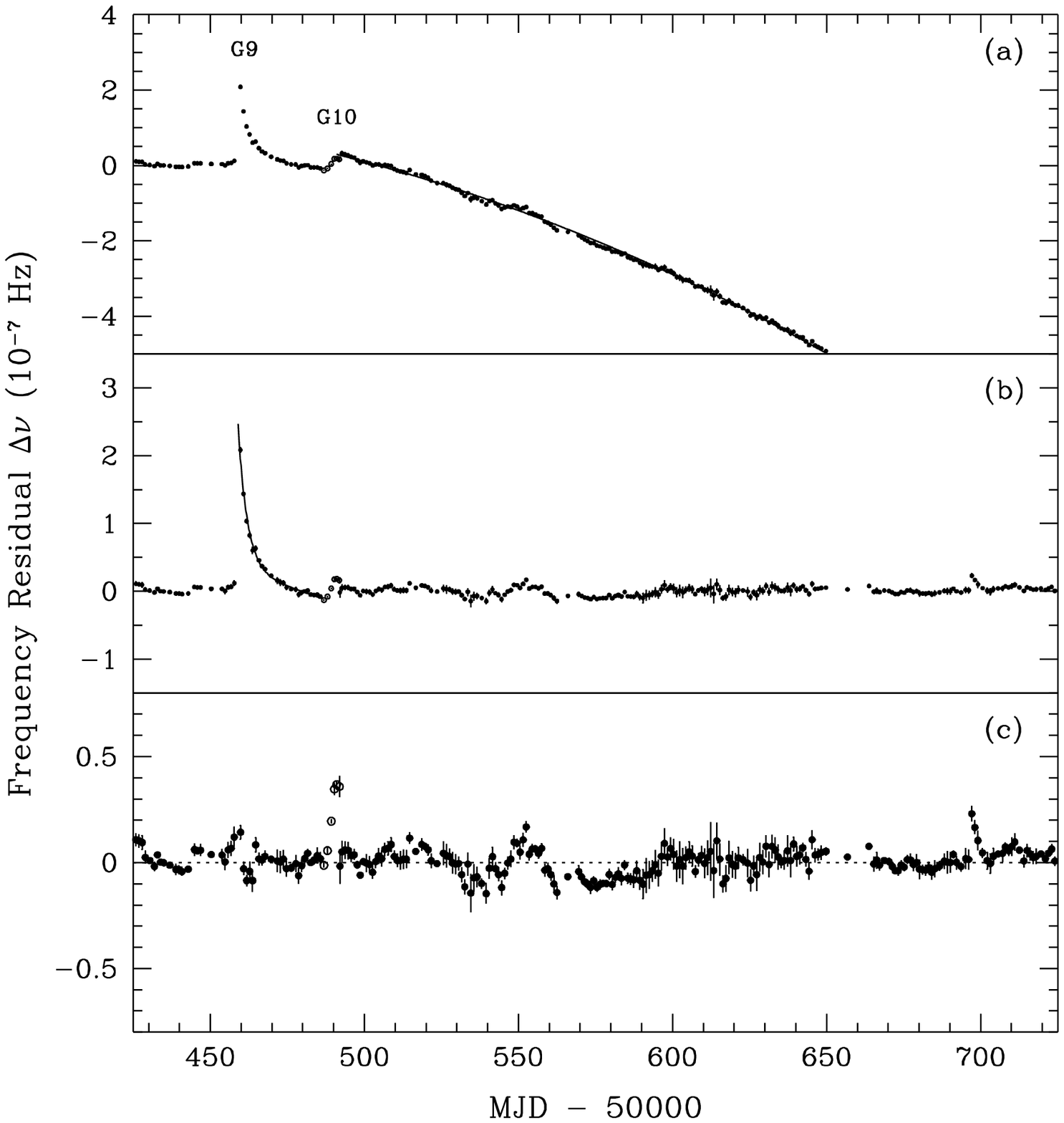}
\figcaption[g9.eps]{
Frequency residuals before and after the early 1997 glitches (G9 and G10),
after subtraction of (a) only the preglitch spindown model; (b) a model
for the February event without a short-term component (Model 10a); (c) the
adopted model for the January event as well (solid line in panel b).
The open circles are JB data that reveal the gradual spinup associated
with Glitch 10.
\label{g9fig}}
\vskip 0.25truein

In some respects the February event resembles the glitches with
gradual spinups observed in 1989 and 1996, yet it lacks the
exponentially decaying term ($\Delta\nu_1$) that dominates in all other
glitches.  Its close proximity in time to the January glitch is also
unusual, and in \S\ref{ashock} we raise the possibility that Glitch 10
is an unusually strong example of an ``aftershock'' phenomenon that may
have been seen, to a much smaller degree, following other glitches as
well.  Although we refer to this event as a glitch, it quite possibly
represents a different type of timing irregularity \citep[see][for a
possible analogue in PSR J1614-5047]{Wang00}.

One difficulty with modeling this pair of glitches is that they occur
just 200~d after the preceding glitch, which leaves only a $\sim$100~d
timespan over which to fit a preglitch model for Glitch 9 if one
wishes to safely exclude transient effects of Glitch 8.  To better
reflect the cumulative effects of the series of glitches, we instead
used the spindown model that was fit to the 500~d prior to Glitch 8
and adjusted its values of $\dot{\nu}$ and $\ddot{\nu}$ by the values
of $\Delta\dot{\nu}_p$ and $\Delta\ddot{\nu}_p$ derived for that
glitch.  This was then used as the preglitch model for Glitch 9, and a
similar procedure was used to create a preglitch model for Glitch 10.
The effects of Glitch 9 are therefore given with respect to Glitch 8,
and Glitch 10 with respect to Glitch 9.

\subsection{DM Jump around MJD 50740\label{dmjump}}

In the middle of 1997 October there was a major disruption of the Crab
pulsar signal which appears to be related to interstellar propagation.
The event was characterized by a sharp decrease in flux as well as a
noticeable jump in the pulse phase.  For approximately a week (October
18--25), the 327 MHz GB data showed a second pulse profile superposed
on the original one and phase shifted by $\sim$0.2 period (6.5 ms);
subsequently the ``new'' profile became dominant and the ``old''
profile faded.  A corresponding phase shift by $\sim$0.08 period (2.5
ms) was seen in the 610 MHz data, although the pulse signal was very
weak during this interval.

%%%%%%%%%%%%%%%%%%%%%%%%%%%%%%%%%%%%%%%%%%%%%%%%%%%%%%%%%%%%%
%%%%%%%%%%%%%%%%%%%%%%%%   FIG. 7   %%%%%%%%%%%%%%%%%%%%%%%%%
%%%%%%%%%%%%%%%%%%%%%%%%%%%%%%%%%%%%%%%%%%%%%%%%%%%%%%%%%%%%%

\vskip 0.25truein
\includegraphics[width=3.3in]{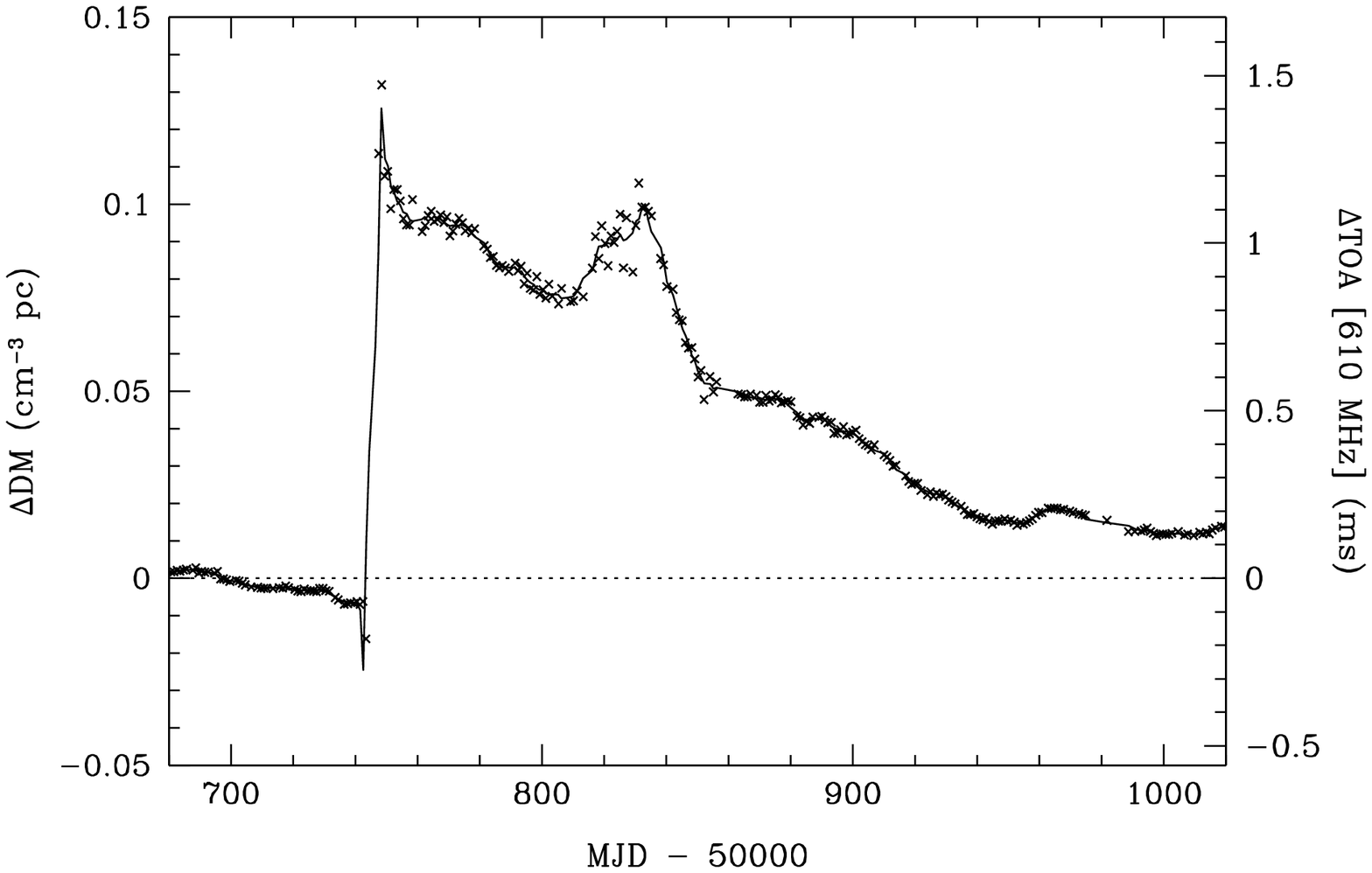}
\figcaption[dmjump.eps]{
DM jump of 1997 October as measured using the 610 MHz GB data and
the scattering-corrected 327 MHz GB data.  The scale on the right-hand
side indicates the expected change in the 610 MHz TOA's resulting from
the DM jump (compare with Figure~\ref{dmerrfig}).  The solid line, a
running 5-day boxcar smooth, is the correction applied to the 610 MHz
data for Glitch 11.
\label{dmj_fig}}
\vskip 0.25truein

The frequency dependence of the shift points to a jump in the
dispersion measure of $\approx 0.15$ pc cm$^{-3}$, which can account
for about 5.8 ms of the shift at 327 MHz and 1.6 ms at 610 MHz.  The DM
jump was followed by a nearly linear recovery towards the original DM
(Figure~\ref{dmj_fig}).  The remaining phase shift appears to have been
non-dispersive in nature.  Since both ``old'' and ``new'' pulses are
seen for several days at 327 MHz, we attribute this shift to changes in
the optical path length due to interstellar refraction, in combination
with timing noise.  More detailed analyses of this event are presented 
in two companion papers (\citealt{Bac00}; \citealt*{Lyn00b}).  We find 
no evidence for a persistent offset in $\nu$ or $\dot{\nu}$ which would 
indicate that the rotation of the star was affected by this event.

\subsection{Glitch 11: MJD 50812\label{g11}}

On 1997 December 30, a rotation glitch occurred that was comparable in
magnitude to Glitch 9 ($\Delta\nu/\nu \sim 9 \times 10^{-9}$). The
decay timescale of the transient, 2.9~d, is also typical of past
glitches.  Phase and frequency residuals for this glitch are shown in
Figure~\ref{g11fig}, relative to a spindown model fit to 300 days
before the glitch (but omitting the period around the October DM
jump).  The residuals have been corrected for the lingering effects of
the DM jump as described by Fig.~\ref{dmj_fig}.  However, as a
comparison between Models 11a and 11b given in Table~\ref{fittbl}
indicates, even the large DM corrections inferred over this period
have little effect on the glitch model, since the DM is not changing
rapidly over the course of the glitch.

%%%%%%%%%%%%%%%%%%%%%%%%%%%%%%%%%%%%%%%%%%%%%%%%%%%%%%%%%%%%%
%%%%%%%%%%%%%%%%%%%%%%%%   FIG. 8   %%%%%%%%%%%%%%%%%%%%%%%%%
%%%%%%%%%%%%%%%%%%%%%%%%%%%%%%%%%%%%%%%%%%%%%%%%%%%%%%%%%%%%%

\vskip 0.25truein
\includegraphics[width=3.3in]{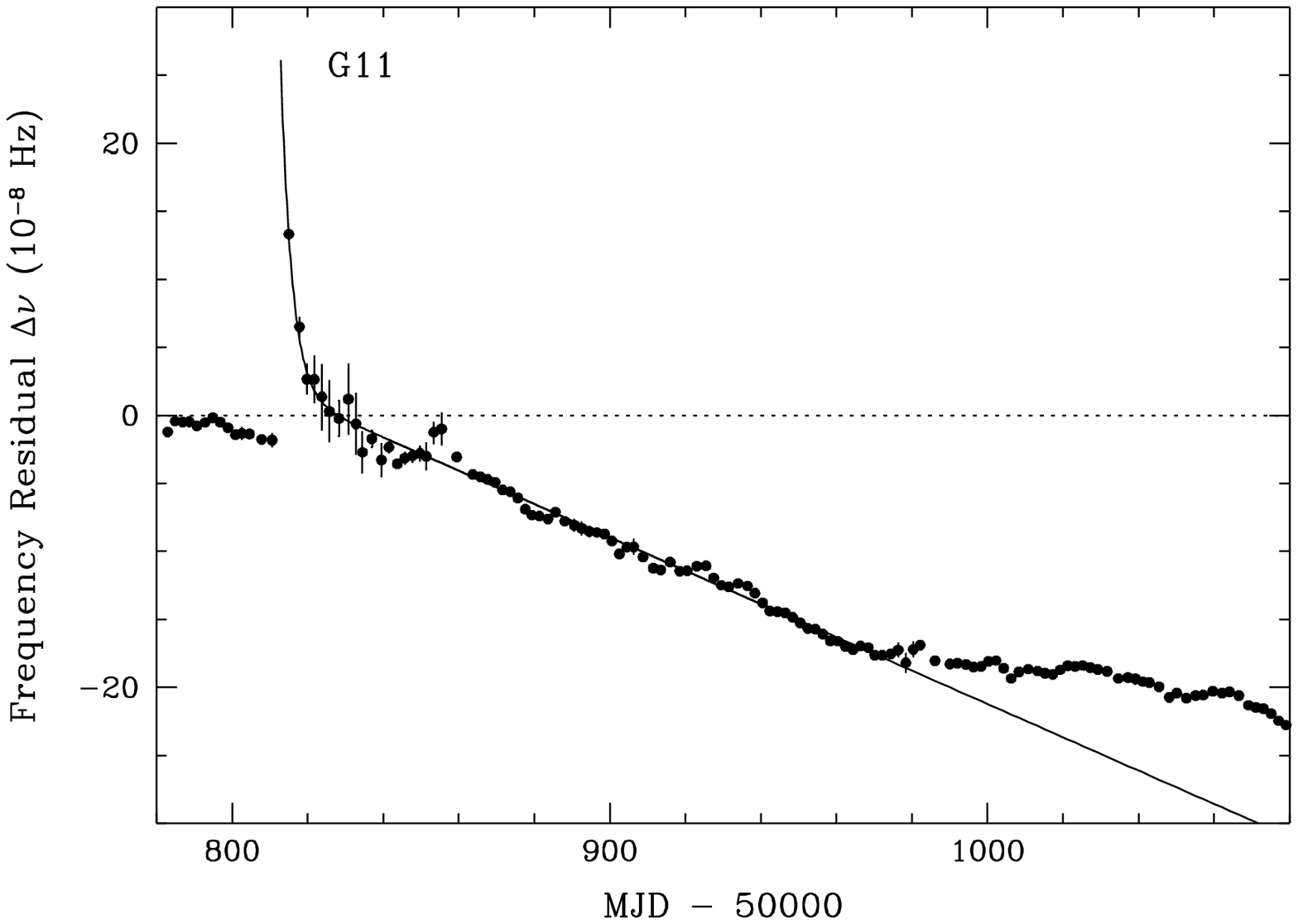}
\figcaption[g11.eps]{
Frequency residuals showing the effects of the small glitch of
1997 December (G11) after applying the DM correction given in
Figure~\ref{dmj_fig}.  The fitted glitch model is shown as the
solid line.
\label{g11fig}}
\vskip 0.25truein

The most striking aspects of the postglitch recovery are a secondary
spinup about 40 days after the glitch (see \S\ref{ashock}) and a
change in $\dot{\nu}$ about 150 days after the glitch, such that the
frequency begins to deviate from the extrapolated $\dot{\nu}$ of the
glitch model (denoted by the solid line in Fig.~\ref{g11fig}).  As
Fig.~\ref{g11fig} indicates, the subsequent value of $\dot{\nu}$ is
close to the preglitch value.  The possibility that
$\Delta\dot{\nu}_p$ at a glitch may not be truly persistent had been
tentatively raised for Glitch 7, but is even more compelling in this
case.  This shift in $\dot{\nu}$ is followed by a period of
particularly strong fluctuations in rotation frequency, as can be seen
in Fig.~\ref{nurecord}.  We attribute these fluctuations to an
increase in timing noise, which may be linked to the high rate of
glitching the pulsar has experienced since 1995.

\subsection{Glitch 12: MJD 51452\label{g12}}

Around MJD 51452 (1999 October 1), yet another glitch occurred of
approximately the same magnitude as the previous one ($\Delta\nu_0/\nu
\sim 10^{-8}$) and with a comparable decay timescale (3.4~d).  We do
not yet have enough data to study the long-term effects of the glitch,
although it does appear to be accompanied by a change in $\dot{\nu}$
as was seen in the previous glitches.  The best-fit parameters to the
90 days following the glitch are shown in Table~\ref{fittbl}, but more
observations will be needed to obtain reliable values of $\Delta\nu_p$
and $\Delta\dot{\nu}_p$.  The preglitch model was fit to 450 days 
before the glitch.

%%%%%%%%%%%%%%%%%%%%%%%%%%%%%%%%%%%%%%%%%%%%%%%%%%%%%%%%%%%%%%%%%%%%%%%%%%%%%

\section{THE FREQUENCY OF CRAB GLITCHES\label{disc_freq}}

In this section we examine the occurrence in time of glitches and
glitch-related events.  Three idealized models are considered: (1)
glitches are spaced evenly in time; (2) glitches are independent events
spaced randomly in time; and (3) the glitch amplitude is proportional
to the time since the last glitch.  Although the statistics are
hampered by the small number of glitches that have occurred during 
continuous observations, inclusion of the recent events appears to favor the
second model (random occurrence).  Finally, we present tentative
evidence that some glitches may be followed by secondary spinups or
``aftershocks.''

\subsection{Intervals between glitches\label{disc_int}}

In order to study the intervals between glitches it is essential that
continuous timing measurements be available.  While very large glitches
in the Crab are expected to leave persistent changes in $\dot{\nu}$
that could be detectable at a later date, it is unlikely that the
smaller events discussed in this paper could be inferred from data
taken months or years afterward.  Hence we restrict our discussion to
glitches since 1982, for which continuous timing measurements from JB
are available.  To determine an appropriate completeness threshold, we
examined a continuous 15-year frequency record (1983--1997) from JB,
based on 3-day averages.  We found that frequency jumps larger than
$\Delta\nu=5\times 10^{-8}$ Hz are clearly identifiable as either
glitches or noise spikes in the data.  Thus all of the glitches
observed since 1986, with the exception of Glitch 10 (which we exclude
from this analysis), should have been detected at any time during this
interval.  This yields a total sample of eight glitches and seven
interglitch intervals.  Although there is some question about whether 
Glitch 7 would have been clearly identified prior to 1995, the results
presented here are qualitatively the same whether it is included or not.

The observed distribution of interglitch intervals between
1983 and 2000 is plotted cumulatively as $S_N(T)$ in
Figure~\ref{cumplt}(a), where $S_N(T)$ is the fraction of glitches
which have preceding intervals $<T$.  The dashed curve represents
the expected distribution for a Poisson process with mean interval
$\lambda$, 
\[P(T) = 1 - e^{-T/\lambda}\,,\]
where $\lambda$ is taken as the mean interglitch interval (684~d).
Despite the small-number statistics, the model distribution appears to
be a fair match to the observations, especially if one considers only
glitches since 1995, which account for all of the shorter intervals
[Fig.~\ref{cumplt}(b)].  A Kolmogorov-Smirnov (K-S) test indicates that
we can only reject the null hypothesis (that the observations are drawn
from a Poisson model distribution) with a probability of $\sim$30\%.
Since the agreement with a Poisson model depends crucially on the more
recent data, continued monitoring of the pulsar's rotation will be
needed to determine whether future glitches continue to follow this
distribution.

%%%%%%%%%%%%%%%%%%%%%%%%%%%%%%%%%%%%%%%%%%%%%%%%%%%%%%%%%%%%%
%%%%%%%%%%%%%%%%%%%%%%%%   FIG. 9   %%%%%%%%%%%%%%%%%%%%%%%%%
%%%%%%%%%%%%%%%%%%%%%%%%%%%%%%%%%%%%%%%%%%%%%%%%%%%%%%%%%%%%%

\vskip 0.25truein
\includegraphics[width=3.3in]{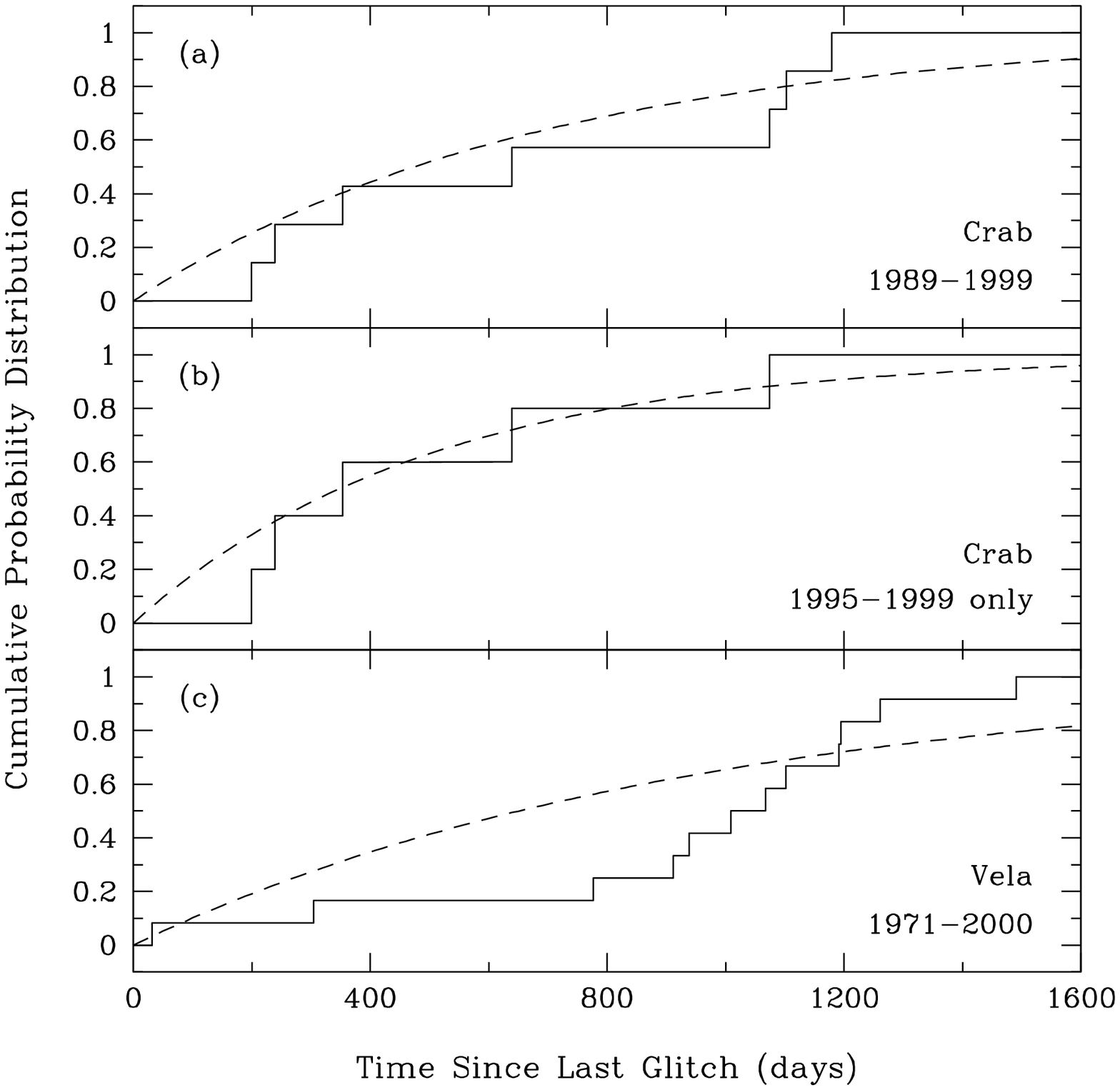}
\figcaption[cumplt.eps]{
Cumulative distribution of interglitch intervals for (a) Crab glitches,
intervals ending in 1989 or later; (b) Crab glitches, intervals ending
in 1995 or later; (c) large Vela glitches, intervals ending in 1971 or
later.  The dashed curves represent models of a Poisson process having
a mean interval given by the average of the intervals in each plot.
\label{cumplt}}
\vskip 0.25truein

If, on the other hand, glitches tended to be spaced evenly in time, one
would expect to see a steep rise in $S_N(T)$ around the characteristic
interglitch interval.  Such a signature is lacking in the Crab but is
present in the Vela pulsar, for which there is a tendency for glitches
to be separated by $\sim$1000~d [Fig.~\ref{cumplt}(c)].  Here we have
included only the thirteen large ($\Delta\nu/\nu > 10^{-7}$) Vela
glitches from 1969--2000; although two much smaller events
($\Delta\nu/\nu \sim 10^{-8}$) have been reported by \citet*{Cor88} and
\citet{Fla95}, they are omitted since it is unclear whether such events
would have been detected over the entire 31-year timespan.  Parameters
for the most recent glitches (in 1996 and 2000) are taken from GB
timing observations of Vela that are concurrent with the Crab
observations.  Although Vela shows a total range of interglitch
intervals that is similar to the Crab, the shape of the distribution is
substantially different, and a K-S test rules out the Poisson model
(with $\lambda$=940~d) at a 96\% confidence level.  Large glitches of
PSR J1341$-$6220 may be even more regular, with four interglitch
intervals of 675$\pm$50~d \citep{Wang00}.  The relatively constant
intervals between Vela glitches have been attributed to a critical lag
between the rotation of the superfluid and crust that must be achieved
in order for a glitch to occur \citep{Alp93}.

In the third of our idealized models, the size of a glitch would be
proportional to the time since the last glitch, for example if glitches
empty an angular momentum reservoir that builds up between glitches at
a roughly constant rate.  Figure~\ref{last} shows that this is
generally {\it not} the case for the Crab: glitches which are separated
by $\sim$1000~d from the previous glitch still vary in amplitude by
over a factor of 20.  (Here the glitch amplitude is defined as
$\Delta\nu_g \equiv \Delta\nu_0 + |\Delta\nu_3|$, i.e.\ any short-term
rise is assumed to be unresolved.)  For the large Vela glitches (solid
circles), there is also no significant correlation, although including
the two small events (squares) does lead to a clear preference for
small glitches to follow shortly after large ones.  Nonetheless, 
the scatter in the relation is large, and the possibility that other
small Vela glitches may have been missed needs to be evaluated.
\citet{Wang00} find that most pulsars exhibit no clear relation between
glitch size and preceding interglitch interval, and conclude that the 
triggering of glitches is not strongly tied to the global spindown of 
the star.

%%%%%%%%%%%%%%%%%%%%%%%%%%%%%%%%%%%%%%%%%%%%%%%%%%%%%%%%%%%%%
%%%%%%%%%%%%%%%%%%%%%%%%   FIG. 10  %%%%%%%%%%%%%%%%%%%%%%%%%
%%%%%%%%%%%%%%%%%%%%%%%%%%%%%%%%%%%%%%%%%%%%%%%%%%%%%%%%%%%%%

\vskip 0.35truein
\includegraphics[width=3.3in]{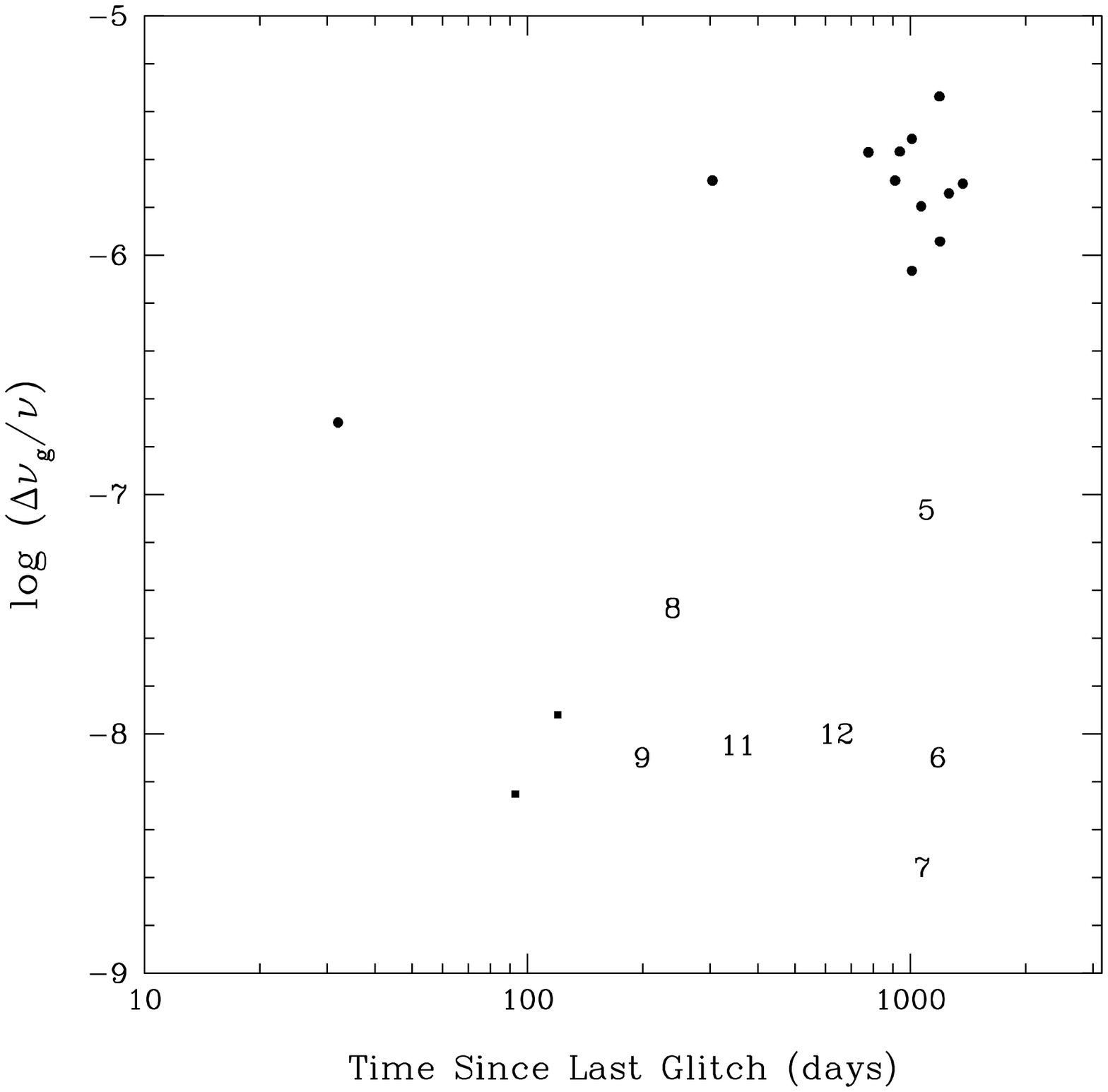}
\figcaption[last.eps]{
Glitch amplitude $\Delta\nu_g/\nu$ plotted against the time since the
previous glitch.  Numbers indicate Crab glitches since 1989
(earlier glitches have been omitted due to possible gaps in the timing
record).  Filled circles indicate major Vela glitches; the two filled
squares are the small Vela glitches that were omitted from
Figure~\ref{cumplt}.
Since they occur soon after large glitches, their inclusion in the glitch
sample leads to only a minor change in the other intervals.
\label{last}}
\vskip 0.35truein

\subsection{Possible Glitch Aftershocks\label{ashock}}

Several of the small glitches observed since 1995 show indications of
secondary spinups occurring $\sim$20--40 days after the main glitch.
These spinups can be identified as the first significant departures
from the smooth exponential decay of the main transient, and generally
show gradual rise times of $\sim$1 day.  The most clear-cut example
follows the 9th glitch, where it is classified as Glitch 10, but other
smaller events (at the $\Delta\nu \sim 10^{-8}$ Hz level) are seen
following Glitches 7 and 11 (Figure~\ref{postfig}).  These latter
events are comparable in magnitude to the timing noise seen between
glitches, and hence their association with the glitch events is
admittedly speculative.  However, their morphological resemblance to
Glitch 10---and the uniqueness of Glitch 10 among well-observed Crab
glitches---suggest that they indeed constitute a separate class of
frequency events.  The deviations from a simple exponential model that
were seen by \citet{Lyn87} in the residuals following Glitch 4 may also
have been the result of an aftershock.

%%%%%%%%%%%%%%%%%%%%%%%%%%%%%%%%%%%%%%%%%%%%%%%%%%%%%%%%%%%%%
%%%%%%%%%%%%%%%%%%%%%%%%   FIG. 11  %%%%%%%%%%%%%%%%%%%%%%%%%
%%%%%%%%%%%%%%%%%%%%%%%%%%%%%%%%%%%%%%%%%%%%%%%%%%%%%%%%%%%%%

\vskip 0.25truein
\includegraphics[width=3.3in]{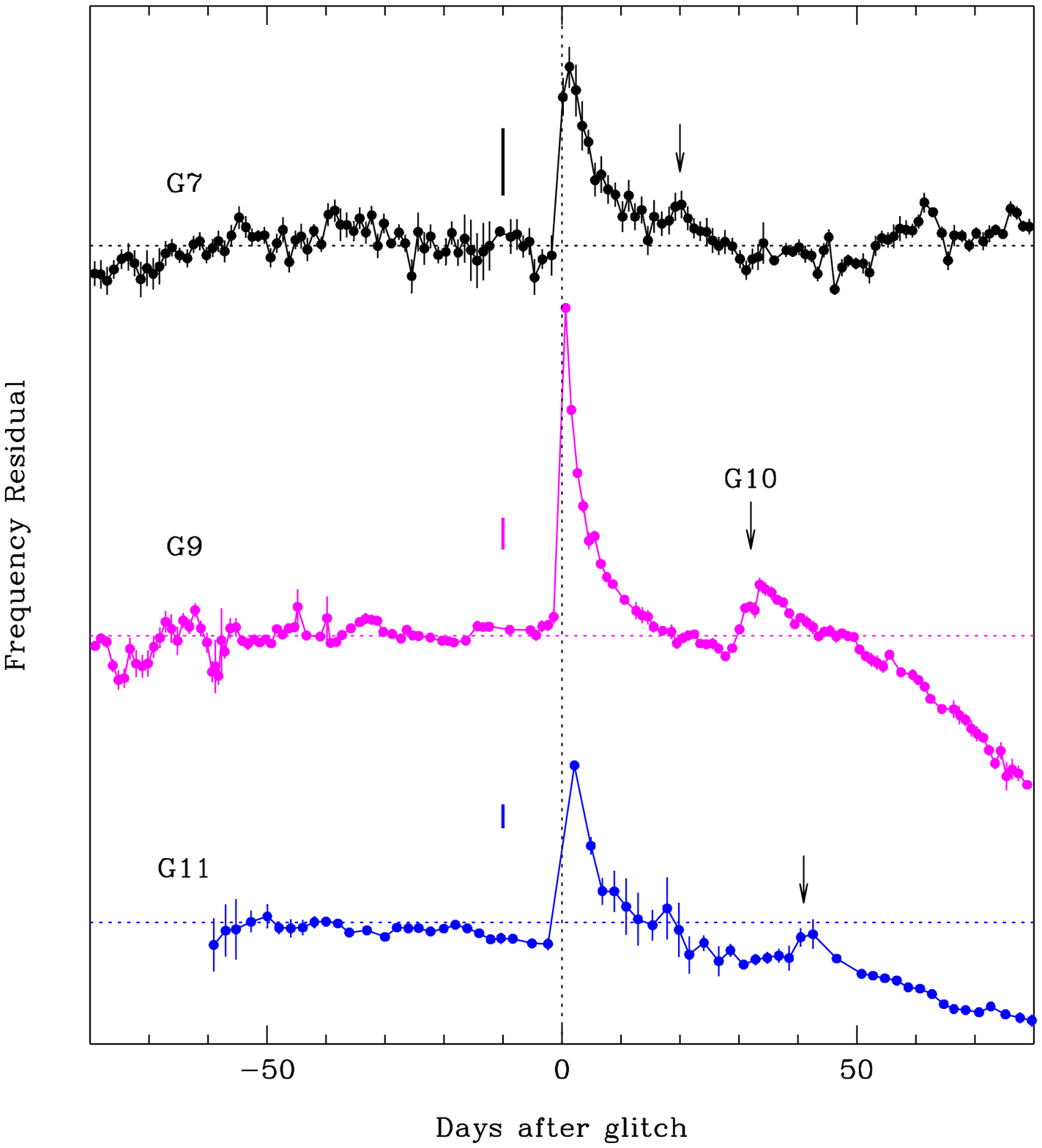}
\figcaption[ashock.eps]{
Frequency residuals for three of the small 1995--1997 glitches.
Times are referenced to the assumed glitch
time, which precedes the first postglitch frequency measurement by up
to a few days.  Arrows indicate times at which secondary spinups occur,
and the vertical bar at Day $(-10)$ corresponds to $\Delta\nu$=$2 \times
10^{-8}$ Hz.
\label{postfig}}
\vskip 0.2truein

Although further observations will be needed to confirm the association
of these spinups with glitches, we suggest that future modeling of
glitches include consideration of the possibility of such aftershocks,
and that efforts be made to search for such features following glitches
in the Crab and other pulsars.  One potential difficulty is that these
events are more apparent following smaller glitches, i.e. they do not
appear to scale with glitch size, making it more difficult to identify
them following large glitches.  Thus, regular monitoring of pulsars
like the Crab which display a variety of glitch sizes will be useful
for exploring this issue.

%%%%%%%%%%%%%%%%%%%%%%%%%%%%%%%%%%%%%%%%%%%%%%%%%%%%%%%%%%%%%%%%%%%%%%%%%%%%%

\section{OBSERVED GLITCH PROPERTIES\label{postg}}

\subsection{Amplitudes of the glitches}\label{disc_ag}

Fig.~\ref{last} displays the well-known difference in the sizes
($\Delta\nu_g/\nu$) of Crab and Vela glitches.  The much larger size
of most Vela glitches, $\sim 10^{-6}$, appears to be characteristic of
at least nine other pulsars as well (\citealt*{Lyn00a};
\citealt{Wang00}).  Still, the Crab is not unique in experiencing
smaller glitches of size $\sim 10^{-8}$; in fact, both Vela and PSR
J1341$-$6220 show glitch sizes spanning a range from
$10^{-8}$--$10^{-6}$ \citep{Wang00}.  What is noteworthy about the
Crab is that it does not experience substantially larger glitches.
Conventionally it has been assumed that this is due to the relative
youth of the Crab pulsar; the higher crustal temperature would allow
stresses due to spindown to be partially relieved by gradual processes
such as plastic flow and vortex creep
\citep[e.g.,][]{McKenna90,Rud91}.  This interpretation is consistent
with the large recovery fractions seen in Crab glitches (\S\ref{nup}).
On the other hand, the recent observation of a large ($\Delta\nu_g/\nu
\sim 6 \times 10^{-7}$) glitch in an anomalous X-ray pulsar
\citep*{Kaspi00} suggests that even hot neutron stars can experience
large glitches.

The same rate of angular momentum transfer from superfluid to crust can
be achieved with frequent small glitches or rare large ones.  Combining
the sizes and frequencies of the glitches, we can define an ``activity
parameter'' $A_g \equiv (\sum \Delta\nu_p)/t_{obs}$, which is the net
angular momentum loss due to glitching over some observing timespan
$t_{obs}$ if persistent changes in $\dot\nu$ are neglected.  The
advantage of $A_g$ as a long-term indicator of glitch effects is that
it is relatively insensitive to the discovery of smaller glitches as
the data quality improves, unlike the analysis in \S\ref{disc_int} for
which a threshold must be explicitly defined.  The main disadvantage is
that $\Delta\nu_p$ is often not well determined for the Crab, since it
depends on the glitch model employed and is generally much smaller than
either the instantaneous frequency jump ($\Delta\nu_0$) or the
frequency change due to the change in $\dot\nu$
($t_{obs}\Delta\dot{\nu}_p$).

%%%%%%%%%%%%%%%%%%%%%%%%%%%%%%%%%%%%%%%%%%%%%%%%%%%%%%%%%%%%%
%%%%%%%%%%%%%%%%%%%%%%%%   FIG. 12  %%%%%%%%%%%%%%%%%%%%%%%%%
%%%%%%%%%%%%%%%%%%%%%%%%%%%%%%%%%%%%%%%%%%%%%%%%%%%%%%%%%%%%%

\vskip 0.2truein
\includegraphics[width=3.3in]{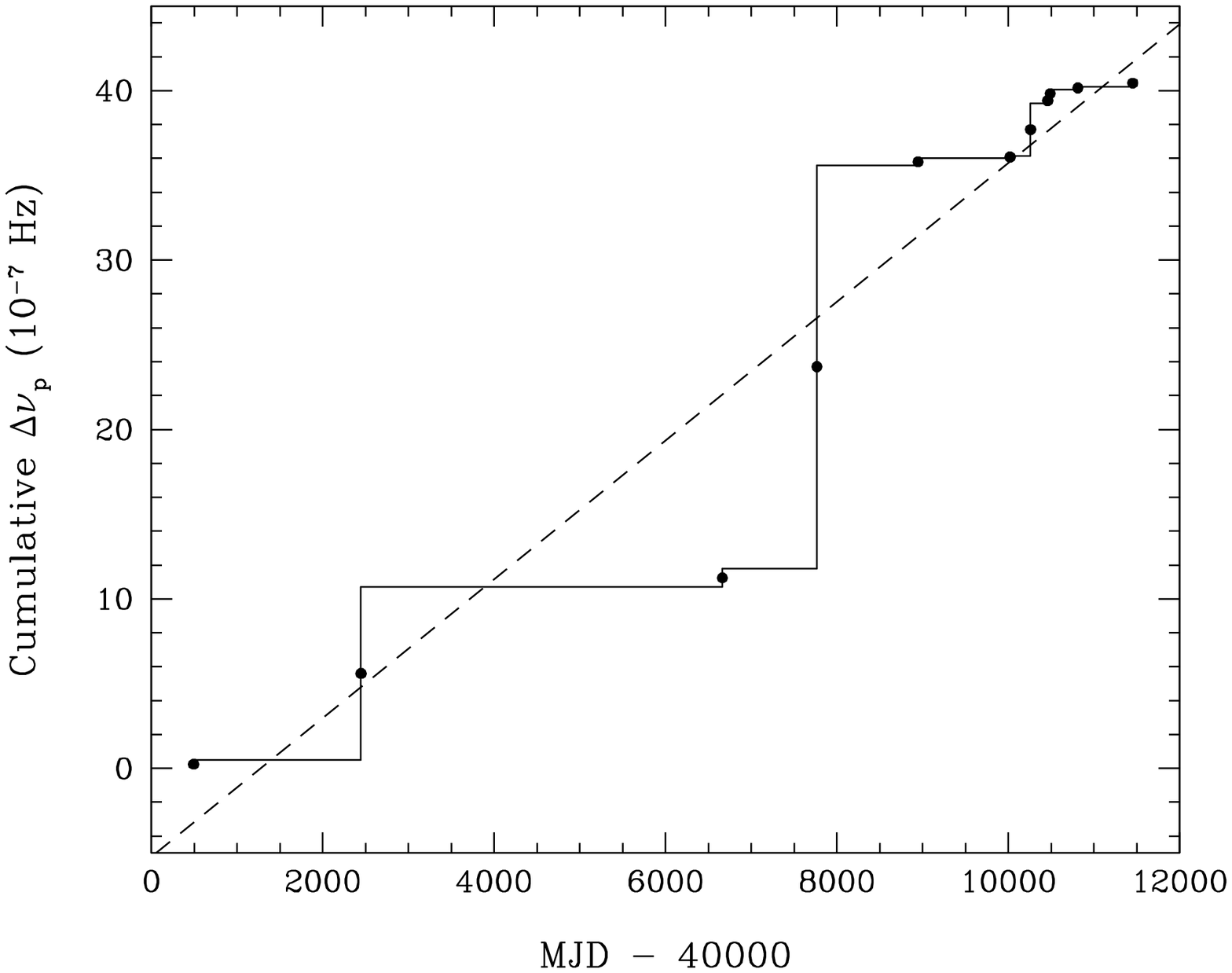}
\figcaption[ag.eps]{
Cumulative $\Delta\nu_p$ due to glitches plotted as a function of time
based on data from Table~\ref{alltbl}.  A least-squares fit to the
midpoints of the frequency jumps is shown as a dashed line.
The slope of this fit provides an estimate of the activity parameter
$A_g$.
\label{ag}}
\vskip 0.2truein

Figure~\ref{ag} shows a cumulative plot of $\Delta\nu_p$ vs.\ time
across the historical glitch record (since 1969).  $A_g$ is given by
the slope of this relation, which is $\sim 1.3 \times
10^{-5}|\dot{\nu}|$ over the entire timespan.  Excluding the recent
series of glitches yields $A_g \sim 1.1 \times 10^{-5}|\dot{\nu}|$, an
insignificant difference given the large scatter around the mean line.
Thus, we see no clear evidence that the rate of angular momentum loss
by the superfluid has significantly increased, despite the more
frequent glitching.

\subsection{Persistent change in $\dot{\nu}$\label{nudot}}

Our observations confirm that large Crab glitches lead to cumulative
increases in $|\dot{\nu}|$ that do not decay in time. The increase per
glitch ranges from $|\Delta\dot{\nu}/\dot{\nu}| \sim 10^{-5}$ to $4
\times 10^{-4}$.  \citet{Alp96} suggest that this is a signature of
the formation of new vortex ``capacitors'' in this young pulsar: regions
that are decoupled from the regular spindown of the star and thus serve
as reservoirs for storing up angular momentum.  As more of the superfluid
decouples, the external torque acts on a (permanently) lower moment
of inertia, increasing $|\dot{\nu}|$.  Recent analysis of glitch
behavior in a large sample of pulsars by \citet{Lyn00a} offers
indirect support for this idea.  In a sample of older pulsars, the
glitch activity parameter $A_g$ (see \S\ref{disc_ag}) was roughly
0.02 $|\dot{\nu}|$, suggesting that $\sim$2\% of the angular momentum
outflow in pulsars is trapped in capacitive regions that only release
their angular momentum in glitches.  However, this percentage is
considerably smaller in the Crab and a few other young pulsars,
consistent with the idea that the youngest pulsars are still in the
process of forming capacitors.

On the other hand, the extremely small value of $A_g/|\dot{\nu}|$ for
the Crab ($\sim$10$^{-5}$) is difficult to reconcile with the
large values of $\Delta\dot{\nu}/\dot{\nu} \sim 10^{-4}$ seen
following glitches.  The cumulative effects of the 1975, 1989, and
1996 glitches already yield $I_{\rm new}/I = \Delta\dot{\nu}/\dot{\nu}
\sim 10^{-3}$ in 25 years, assuming no change in the external torque.
Extrapolating this to the lifetime of the Crab ($10^3$ yr) would imply
that $\sim$4\% of the star's moment of inertia has already been
converted into capacitors, over three orders of magnitude larger than
the fraction of angular momentum released in glitches,
$A_g/|\dot{\nu}|$.  If this is in fact the case, then most of the
excess angular momentum accumulating in the new capacitors cannot be
released by glitching, but must continue to build over time.

Alternatively, the change in $\dot{\nu}$ may reflect a change in the
external torque, perhaps due to a change in the angle between the
rotation and magnetic axes \citep*{Lin98} or an increase in the dipole
magnetic field \citep*{Rud98}.  If the effect is due to a change in the
misalignment angle $\alpha$, then some mechanism must ensure that
glitches only cause $\alpha$ to increase in young pulsars such as the
Crab.  The inferred rate of change of $\alpha$ is $\sim 1.5 \times
10^{-5}$ rad yr$^{-1}$, small enough to satisfy current observational
constraints.

%%%%%%%%%%%%%%%%%%%%%%%%%%%%%%%%%%%%%%%%%%%%%%%%%%%%%%%%%%%%%
%%%%%%%%%%%%%%%%%%%%%%%%   FIG. 13  %%%%%%%%%%%%%%%%%%%%%%%%%
%%%%%%%%%%%%%%%%%%%%%%%%%%%%%%%%%%%%%%%%%%%%%%%%%%%%%%%%%%%%%

\vskip 0.2truein
\includegraphics[width=3.3in]{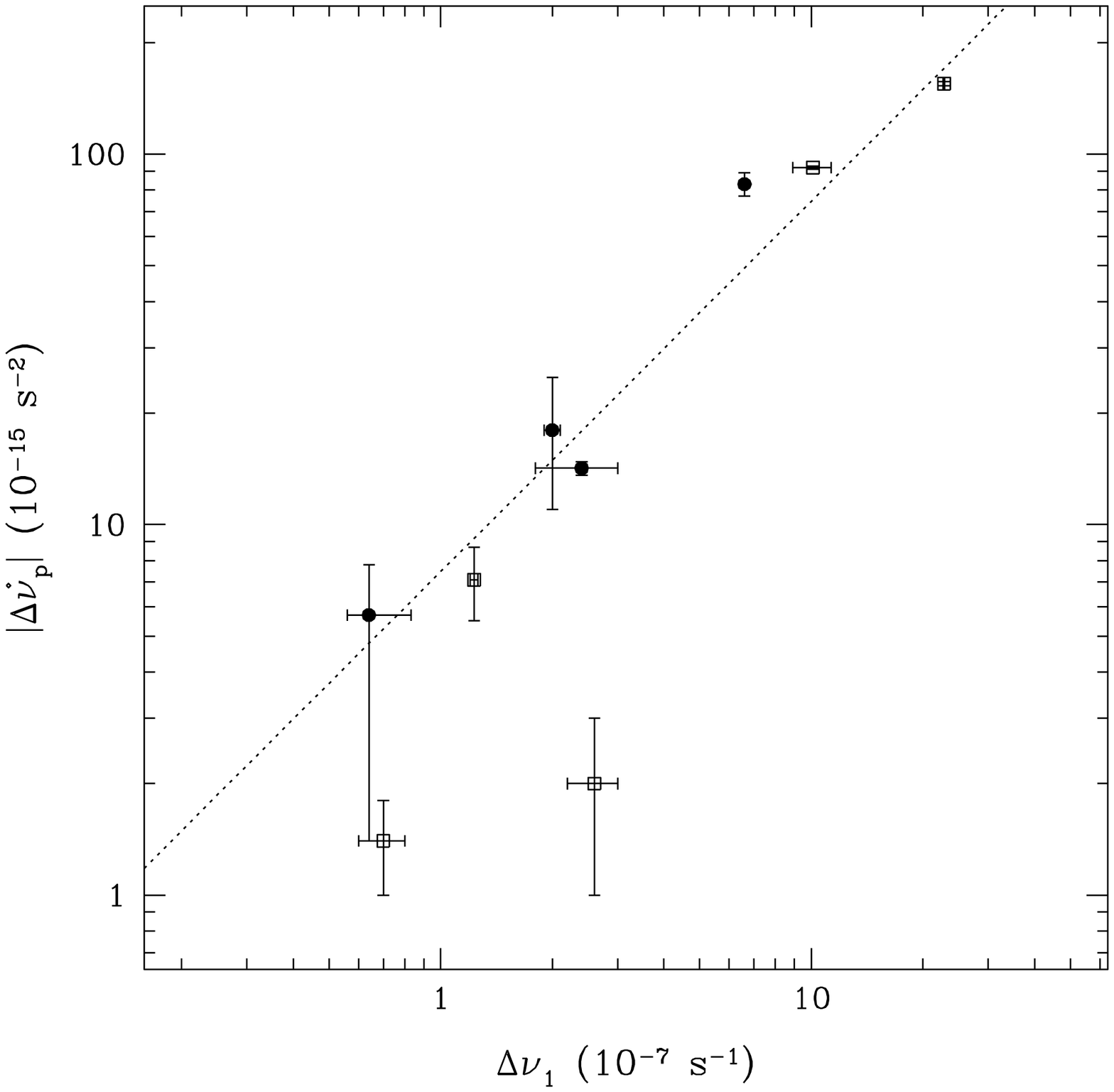} 
\figcaption[nudotlg.eps]{
Observed correlation between the transient jump $\Delta\nu_1$ and the
persistent shift in spindown rate $\Delta\dot{\nu}_p$.  The dotted
line has a slope of $7.5 \times 10^{-8}$ s$^{-1}$ or 1/(150 d).  Open
squares denote events before 1995 and filled circles are events
presented in this paper.  The most recent glitch (Glitch 12) has been
omitted since its long-term recovery is not well sampled.
\label{nudotcor}}
\vskip 0.1truein

Although it is unclear which (if any) of these models provides the
correct interpretation of the $\Delta\dot{\nu}_p$ term, the ensemble of
Crab data provides two important constraints.  First of all, as shown
in Figure~\ref{nudotcor}, the change in $\dot{\nu}$ at a glitch shows a
good correlation with the transient jump $\Delta\nu_1$, with a linear
correlation coefficient of 0.96 (a Spearman rank correlation gives a
coefficient of 0.78).  The only strongly discrepant points (at the
bottom) correspond to the 1969 glitch, whose errors may well be
underestimated, and the 1992 glitch.  Since $\Delta\nu_1$ dominates the
frequency jump for most Crab glitches, $\Delta\dot{\nu}_p$ also
correlates well with the initial jump $\Delta\nu_0$; it shows a
somewhat weaker correlation with $\Delta\nu_p$, but displays no
correlation with the time since the previous glitch (see also
\S\ref{disc_int}).  These results indicate that the $\dot{\nu}$ change
is closely tied to the glitch process, and is not due to some unrelated
process that is waiting for a trigger.  Secondly, we have found that in
Glitch 11 and possibly Glitch 7, the change in $\dot{\nu}$ following
the glitch is not truly persistent, although it does persist for much
longer than the exponentially decaying term.  This suggests that
whatever structural changes lead to a long-term increase in $|\dot\nu|$
can be partially undone at some later point, particularly when the jump
in $\dot\nu$ is relatively small.

\subsection{Decay time of principal transient\label{tau}}

The jump in frequency at a glitch is followed by a partial relaxation
back towards the original frequency, which is generally interpreted as
the re-establishment of an equilibrium lag between the superfluid and
the crust.  In this picture, there exist ``resistive'' regions where a
continous vortex current transfers angular momentum from the
superfluid to the crust, alongside the capacitive regions where the
superfluid is disconnected from the spindown of the crust.  Thus, in
resistive regions the crust normally feels both a {\it decelerative}
torque from the magnetic field and an {\it accelerative} torque from
the faster-rotating superfluid.  However, when the lag is suddenly
reduced in a glitch (due to excessive unpinning of vortices), the
accelerative torque is also reduced and the crust spins down more
rapidly ($|\dot{\nu}|$ increases) until equilibrium is restored by
vortex repinning.  In the vortex creep model of \citet{Alp93}, the
recovery is exponential in time, with a timescale that reflects the
ratio of the pinning energy $E_p$ to the temperature $kT$.

As shown in Figure~\ref{taucorfig}, the main transient component
($\Delta\nu_1$), observed in 10 of the 12 glitches, 
has a recovery timescale that varies from about 3 to 18
days.  Of the glitches since 1995, 4 have recovery times of $\sim$3~d
and one (the largest) has a recovery time of $\sim$10~d.  
Although there is no strong correlation between the glitch
amplitude and the recovery timescale, there is a slight tendency for
larger glitches to have longer recovery times.  One must be cautious,
however, in comparing data from different timing programs, since
differences in time sampling, the length of time used for the fit, or
the model fitting methods employed may have significant effects on the
inferred recovery timescales.  For this reason we have purposely omitted
the 1969 glitch, for which the exponential timescale is highly
uncertain \citep{Boy72}.  In light of these concerns,
probably the strongest conclusion that can be drawn from
Fig.~\ref{taucorfig} is the absence of large glitches with short
recovery timescales.  This is consistent with theories in which $\tau$
increases with pinning strength (e.g.\ the vortex creep model), since
larger glitches would tend to result when larger pinning energies were
overcome.

%%%%%%%%%%%%%%%%%%%%%%%%%%%%%%%%%%%%%%%%%%%%%%%%%%%%%%%%%%%%%
%%%%%%%%%%%%%%%%%%%%%%%%   FIG. 14  %%%%%%%%%%%%%%%%%%%%%%%%%
%%%%%%%%%%%%%%%%%%%%%%%%%%%%%%%%%%%%%%%%%%%%%%%%%%%%%%%%%%%%%

\vskip 0.2truein
\includegraphics[width=3.3in]{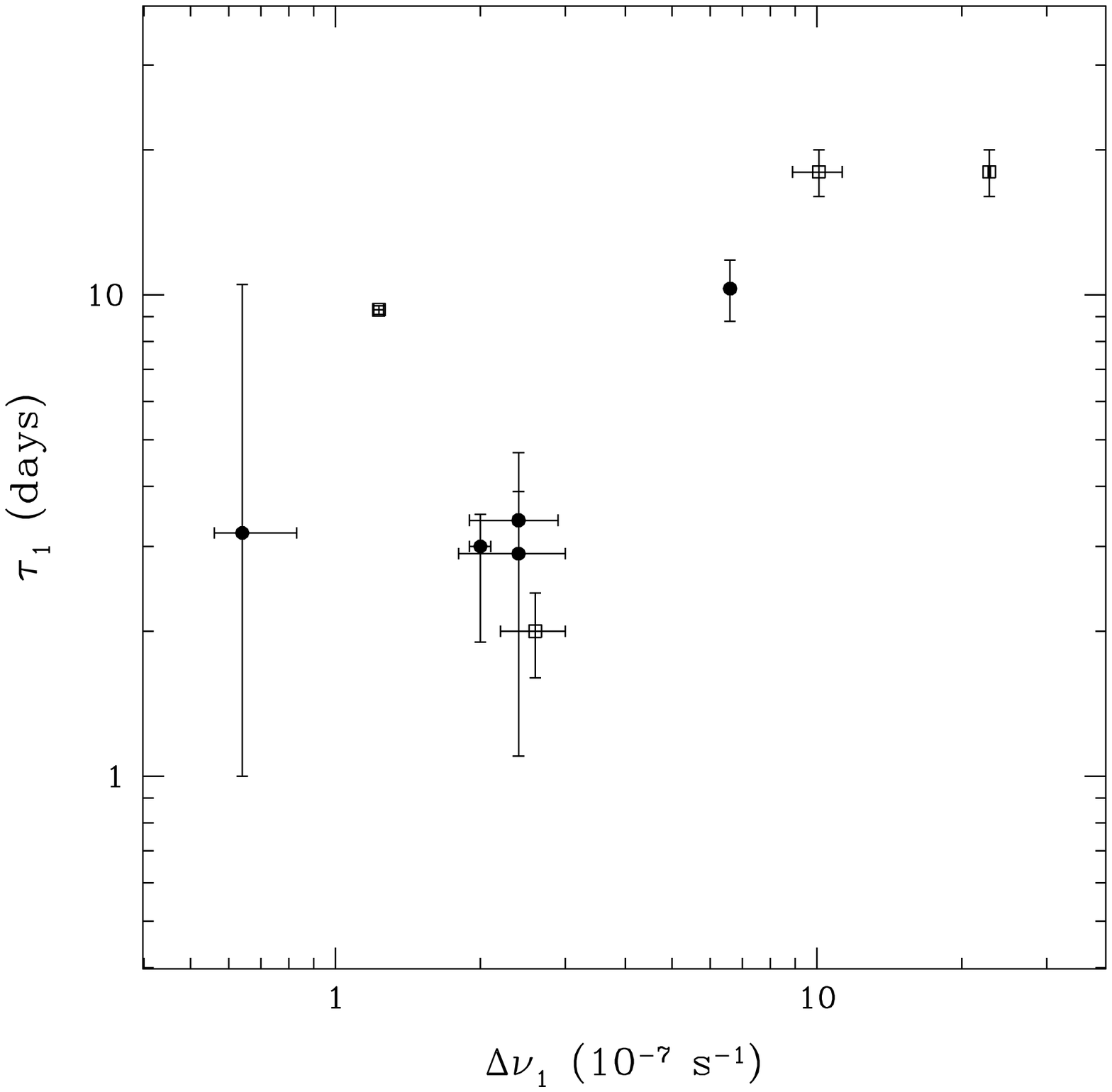} 
\figcaption[taulg.eps]{
Comparison between the transient frequency jump $\Delta\nu_1$
and its decay timescale $\tau_1$.  Open squares denote events before
1995 and filled circles are events presented in this paper.  Glitch
1 has been omitted due to poor time sampling after the glitch.
\label{taucorfig}}
\vskip 0.2truein

In comparison to the Crab, the recovery times following Vela glitches
appear to be extremely regular.  Analyzing the postglitch relaxation of
the first nine glitches observed in that pulsar, \citet{Chau93} found
consistent timescales of $3.2 \pm 0.2$ and $33 \pm 4$ d, independent of
the glitch size.  (Note that these timescales are observed {\it
simultaneously}, unlike in the Crab where a single exponential decay
typically dominates each glitch.)  For the 1988 and 1991 Vela glitches,
an additional component of $10 \pm 3$ hours was resolved.  Fitting a
glitch model to our GB data over the 100~d following the recent Vela
glitch of 1996 October yields similar timescales of 3.0 and 32.0 d.
Thus, our data do not appear to support the suggestion by 
\citet{Alp96} that Crab recovery timescales can be scaled by a constant 
factor to those seen in Vela.

\subsection{Glitch recovery fraction\label{nup}}

As noted in \S\ref{nudot} and \S\ref{tau}, the vortex creep theory of
\citet{Alp93} postulates the existence of two types of pinning regions
in the superfluid, ``capacitive'' regions that transfer angular
momentum to the crust only in glitches, and ``resistive'' regions that
transfer angular momentum continuously but maintain a faster
equilibrium rotation than the crust.  The capacitive and resistive
regions give rise, respectively, to the persistent and decaying jumps
in frequency that are observed at glitches.  Thus, the recovery fraction
\begin{equation}
Q \equiv 1 - \frac{\Delta\nu_p}{\Delta\nu_0}
\end{equation}
indicates the relative importance of resistive and capacitive regions
in postglitch recovery.  \citet{Lyn00a} show that $Q$ is a strong and
monotonic function of $\dot{\nu}$: young pulsars like the Crab show
nearly complete recovery, while older pulsars show almost no recovery.
This observation suggests that as pulsars age, resistive regions give
way to capacitive ones.

An accurate test of this prediction using our data is hampered by
several difficulties in assigning reliable values for $Q$ to Crab
glitches.  First of all, $Q$ is dependent on the model fit:  the
permanent frequency change $\Delta\nu_p$ can only be determined once
the contributions of all transient terms and the change in $\dot{\nu}$
have been removed, and there is no guarantee that we are properly
modeling these effects.  Also, the initial jump $\Delta\nu_0$ is
model-dependent, derived from an extrapolation of the exponential model
back to the time of the glitch.  Finally, it is unclear what part of
the spinup actually recovers: just the unresolved frequency jump
$\Delta\nu_0$, or a combination of $\Delta\nu_0$ and any short-term
transient $\Delta\nu_3$.  The fact that the main transient
$\Delta\nu_1$ exceeds $\Delta\nu_0$ for the 1989 glitch supports the
latter interpretation, which we adopt here.

In light of these uncertainties, we define for simplicity the parameter
\begin{equation}
\tilde{Q} \equiv \frac{\Delta\nu_1}{\Delta\nu_0 + |\Delta\nu_3|}\,,
\end{equation}
which we use as a rough estimate of the recovery fraction.  Values of
this parameter for Crab glitches where it is well-defined are given in
Table~\ref{alltbl}; it is clear that when measurable, the recovery
fraction is large ($\sim$90\%).  In the framework of the vortex creep
model, this is consistent with the interpretation that the Crab is
dominated by resistive rather than capacitive regions of vortex pinning
(\S\ref{nudot}).

\subsection{Time-resolved spinups\label{nu3}}

A portion of the initial spinup in Crab glitches can take much longer
than the core-crust coupling timescale of $<1$ minute.  In the 1989
glitch, part of the spinup was resolved with a 0.8 day timescale
\citep{LSP92}, and we have noted evidence for a $\sim$0.5 day timescale
in the 1996 glitch (\S\ref{g8}).  On the other hand, the Vela glitch
of 1988 December exhibited a spinup that was complete in less than 2
minutes \citep{Mcc90}.  The gradual spinups in the Crab have been
linked by \citet{Alp96} to the formation of vortex traps: during this
process, some vortex lines move in toward the rotation axis,
counteracting the transfer of angular momentum to the crust and
resulting in an extended spin-up.  In the model of \citet{Lin96}, on
the other hand, glitches are induced by injection of heat into the
inner crust (e.g., by starquakes), and the duration of the spinup is
inversely related to the strength of the glitch.  This is because the
superfluid-lattice coupling is highly sensitive to temperature, so
greater heating leads to faster and stronger glitches.  Finally,
\citet{Rud98} suggest that resistance of magnetic flux tubes in the
core to inward motion of vortex lines after the glitch may lead to a
delayed spinup.

Our inability to resolve spinups in most of the smaller Crab
glitches---especially the 1997 January event, for which observations
commenced $\sim$1 hr after the glitch---argues against models in which
smaller glitches occur more gradually \citep[e.g.,][]{Lin96}.  In fact,
excluding Glitch 10, which is unusual in other respects, one is
led to the conclusion that only the largest glitches will have
observable rise times.  Such a trend is consistent with the longer rise time
seen in 1989 as compared to 1996.  While the much larger yet extremely
rapid spinups of the Vela pulsar do not seem to fit within this
framework, age-related differences between the pulsars (e.g., in the
internal temperature) may play a role.

\subsection{Long-term asymptotic rise\label{nu2}}

A long-term asymptotic rise, such as that seen after the 1989 glitch,
has not been unambiguously seen in any of the more recent glitches,
possibly because the exponential timescale involved exceeds the
interval between glitches.  Consequently, some portion of the inferred
``permanent'' jumps in $\dot{\nu}$ and $\ddot{\nu}$ may actually be due
to a long-term exponential.  Indirect evidence for this interpretation
comes from the fact that the braking index
$n\equiv\nu\ddot{\nu}/\dot{\nu}^2$, when measured over intervals
well-separated from glitches, shows a remarkably constant value
\citep[2.51 $\pm$ 0.01,][]{LPS93}, whereas the changes in $\ddot{\nu}$
given in Table~\ref{fittbl} would lead to changes in $n$ of
10\%--15\%.  However, as discussed in \S\ref{results}, an asymptotic
exponential has the wrong curvature for explaining the residuals after
Glitch 8, and provides a poor fit to the residuals after Glitch 10.
Future timing observations during a long ``quiescent'' period in glitch
activity would provide better constraints on the long-term glitch
recovery process.

%%%%%%%%%%%%%%%%%%%%%%%%%%%%%%%%%%%%%%%%%%%%%%%%%%%%%%%%%%%%%%%%%%%%%%%%%%%%%

\section{SUMMARY AND CONCLUSIONS\label{conc}}

We have presented new timing observations of the Crab pulsar at 610 MHz
during a period of increased glitching.  The occurrence of 6 rotation
glitches over a period of four years, of which 4--5 were large enough
to have been detected by previous monitoring efforts, marks a departure
from the relatively long intervals (3--6 years) between glitches
reported prior to this period.  As a result, the range of interglitch
intervals has widened considerably, and shows a distribution that is
consistent with a random (Poisson) process.  Continued monitoring of
this young pulsar will be valuable in determining whether the occurence
of glitches is primarily a regular or stochastic process.

We have also fitted simple glitch models to the data to better
characterize the basic properties of Crab glitches.  We
summarize these properties as follows:

\begin{enumerate}

\item No correlation is found between the glitch amplitude and the time
since the previous glitch.  If vortex pinning occurs at a constant
rate, this implies that glitches do not lead to the complete unpinning
of vortices in a certain region.

\item In agreement with previous studies \citep[e.g.,][]{LPS93}, we
find that Crab glitches tend to be accompanied by long-lasting (in most
cases permanent) changes in spindown rate $\dot\nu$.  This change in
$\dot\nu$ is correlates well with the transient jump $\Delta\nu_1$ or
the glitch amplitude $\Delta\nu_0$, and was observed to vanish or fade
$\sim$50--100 days after two of the smallest glitches.

\item The glitch amplitudes span a range of over an order of magnitude,
from $\Delta\nu/\nu \sim 2 \times 10^{-9}$ to $3 \times 10^{-8}$.
Smaller events are unlikely to be detectable in many cases due to the
substantial timing noise in this pulsar.

\item The average rate of spindown due to glitches (given by the
activity parameter $A_g$) has not changed significantly in recent
years, despite the higher rate of glitching.  This is a result of the
relatively small persistent frequency jumps $\Delta\nu_p$ accompanying
the recent glitches.

\item Although a strong correlation between glitch amplitude and 
recovery timescale is not found, there appears to be an absence of large 
glitches with short recovery times, consistent with the hypothesis that 
the recovery time increases with the vortex pinning energy.

\item The large 1996 glitch exhibits a gradual spinup similar to
that observed in the giant 1989 glitch, with a slightly shorter
timescale (0.5~d rather than 0.8~d).  The next glitch was a 
factor of $\sim$3 smaller and completed its spinup in under 1 hour,
contradicting the hypothesis that smaller glitches have more
gradual spinups \citep{Lin96}.  

\item Detailed observations of the postglitch recovery suggest
secondary spinups or ``aftershocks'' may follow 20--40 d after each
glitch, although these events cannot clearly be distinguished from 
timing noise except in the case of Glitch 10 (1997 February).

\item Two of the recent glitches (1996 June and 1997 February) are
accompanied by substantial changes in the frequency second derivative
$\ddot\nu$, which would imply a change in the braking index $n$ by 10--20\%
from its fiducial value of 2.51.  A more likely explanation is that 
these are due to long-term transients associated with the glitches, 
as was seen following the 1989 event, that were interrupted by 
further glitches.  

\end{enumerate}

Our observations confirm that important differences exist between
glitches in the Crab and Vela pulsars, as has been noted by other
recent studies of pulsar glitches \citep{Lyn00a,Wang00}.  Vela glitches
occur at more or less regular intervals of 2--3 years, and nearly all
possess similar amplitudes and recovery timescales.  A spinup in Vela
has never been resolved in time and the recovery fraction is fairly
small ($\sim$20\%, \citealt*{LPS95}).  All of these properties are at
odds with what has been observed in the Crab.  Furthermore, most Crab
glitches display persistent increases in $|\dot\nu|$, and no Crab
glitches have been observed with sizes $\Delta\nu_g/\nu \sim 10^{-6}$,
whereas such ``giant'' glitches have been seen in at least five young
pulsars with characteristic ages $\tau_c \equiv \nu/(2|\dot\nu|)$ = 
7--16 kyr \citep{Wang00}, including Vela.  If these differences in glitch
behavior are primarily a result of evolution, they imply substantial
structural changes occurring within neutron stars over their first 10 kyr.

\acknowledgments

We thank R. S. Pritchard for help in compiling the Jodrell Bank timing
data.  We are grateful to the NRAO staff for maintenance of the 85ft
Pulsar Monitoring Telescope at Green Bank and for assistance with our
observational program.  Maintenance and operations of the telescope
have also been supported through Naval Research Laboratory and U.S.
Naval Observatory activities with the Green Bank 85ft telescopes.  We
thank D. Nice for his early contributions to the Green Bank effort,
A. Melatos for stimulating discussions about glitch models, and
A. Somer for comments on an earlier draft of this paper.  Helpful
comments from the referee helped to improve the paper substantially.
Pulsar research at Berkeley has been supported by the University of
California and NSF grants AST-9307913 in the past and currently
AST-9820662.  Pulsar research at Jodrell Bank is supported by a grant 
from the UK Particle Physics and Astronomy Research Council.

\bibliographystyle{apj}
\bibliography{glitch}

\clearpage

\begin{deluxetable}{lr}
\tablewidth{0pt}
\tablecaption{Crab timing parameters prior to 1996 glitch.\label{paramtbl}}
\tablehead{\colhead{Parameter} & \colhead{Value}}
\startdata
Right Ascension (J2000) & $05^h34^m31.\!\!^s973$ \\
Declination (J2000) & $+22^\circ 00^\prime 52$\farcs07 \\
Timespan for fit (MJD) & 49759--50259 \\
$\nu$ (s$^{-1}$) & 29.887774244 \\
$\dot{\nu}$ (s$^{-2}$) & $-3.7569761305 \times 10^{-10}$ \\
$\ddot{\nu}$ (s$^{-3}$) & $1.1857849321 \times 10^{-20}$ \\
Braking index $n$=$\ddot{\nu}\nu/\dot{\nu}^2$ & 2.51 \\
\enddata
\tablecomments{Astrometric coordinates are taken from \citealt*{LPS93}.}
\end{deluxetable}

\begin{deluxetable}{llclrrrccrc}
\tablewidth{0pt}
\tabletypesize{\scriptsize}
\renewcommand{\arraystretch}{1.5}
%\rotate
\tablecaption{Model fits to the 1995--1999 Crab glitches.\label{fittbl}}
\tablehead{
	& \colhead{$t_g$} 
	& & \colhead{$\Delta\nu_n$} & \colhead{$\tau_n$}
	& \colhead{$\Delta\nu_p$}
	& \colhead{$\Delta\dot{\nu}_p$}
	& \colhead{$\Delta\ddot{\nu}_p$} \\
	\colhead{Model} & \colhead{(MJD)} 
	& \colhead{$n$} & \colhead{($10^{-7}$ Hz)} & \colhead{(days)}
	& \colhead{($10^{-7}$ Hz)} 
	& \colhead{($10^{-15}$ s$^{-2}$)} 
	& \colhead{($10^{-21}$ s$^{-3}$)} 
	& \colhead{$\tilde\chi^2(\phi_m')$} 
	& \colhead{$\tilde\chi^2(\nu_m')$} 
	& \colhead{Comments} }
\startdata
\phn 7a* & 50020.6 \phn & 1 & \phs\phn 0.64 & 3.2 
	& 0.15 \phn &  $-5.7$ \phn\phn & \nodata & 7.4 & 2.5 \phn\phn
	& Fit to 50020--50070 \\*
\phn 7b  & 50020.6 \phn & 1 & \phs\phn 0.67 & 6.3 
	& \nodata \phn & \nodata \phn\phn & \nodata & 66. & 6.4 \phn\phn
	& Transient only \\
\tableline
\phn 8a  & 50260.07 & 1 & \phs\phn 6.5 \phn & 10.4 
	& 3.12 \phn & $-82.9$ \phn\phn & \phs 0.85 & 831. & 5.0 \phn\phn
	& Fit to 50260--50450 \\*
\phn 8b  & 50260.07 & 1 & \phs\phn 6.4 \phn & 10.2 
	& 1.23 \phn & $-66.7$ \phn\phn & \nodata & 911. & 5.0 \phn\phn
	& No $\Delta\ddot{\nu}_p$ term, \\*
   & & 2 & \phs\phn 2.0 \phn & 111.\phn & & & & & & slow expon.\ decay \\
\phn 8c* & 50259.93 & 1 & \phs\phn 6.6 \phn & 10.3 
	& 3.14 \phn & $-83.0$ \phn\phn & \phs 0.86 & 833. & 4.5 \phn\phn
	& Short-term rise \\
   & & 3 & \phn $-3.1$ \phn & 0.5 \\
\tableline
\phn 9* & 50459.15 \phn & 1 & \phs\phn 2.01 & 3.0 
	& 0.32 \phn & $-18.2$ \phn\phn & \nodata & 8.8 & 2.8 \phn\phn
	& Fit to 50459--50487 \\
\tableline
10a & 50489 \phd\phn & \nodata & \phn\nodata & \nodata 
	& 0.47 \phn & $-4.4$ \phn\phn & $-2.15$ & 2215. & 11.7 \phn\phn
	& Fit to 50491--50811 \\
10b* & 50489 \phd\phn & 3 & \phn $-0.33$ & 2.2 
	& 0.50 \phn & $-4.8$ \phn\phn & $-2.13$ & 3004. & 8.9 \phn\phn
	& Incl.\ JB 50486-91 \\
10c & 50489 \phd\phn & 2 & $-33.3$ & 381.\phn
	& 33.7 \phn\phn & $-101.4$ \phn\phn & \nodata & 10927. & 21.9 \phn\phn
	& Asymptotic expon. \\*
\tableline
11a & 50812.9 \phn & 1 & \phs\phn 2.46 & 2.5 
	& 0.22 \phn & $-14.5$ \phn\phn & \nodata & 36.9 & 5.7 \phn\phn
	& Fit to 50812--50950 \\*
11b* & 50812.9 \phn & 1 & \phs\phn 2.44 & 2.9 
	& 0.17 \phn & $-14.2$ \phn\phn & \nodata & 38.1 & 4.9 \phn\phn
	& DM-corrected \\*
\tableline
12* & 51452.3 \phn & 1 & \phs\phn 2.43 & 3.4 
	& 0.44 \phn & $-6.0$ \phn\phn & \nodata & 770. & 39.1 \phn\phn
	& Fit to 51453--51543 \\*
\enddata
\tablecomments{Each transient component is classified as a short to
intermediate term decay ($n$=1), a long-term rise or decay ($n$=2), or
a short-term asymptotic rise ($n$=3). Asterisks (*) denote adopted fits.} 
\end{deluxetable}

\clearpage

\begin{deluxetable}{ccclcllllcc}
\tabletypesize{\scriptsize}
%\rotate
\tablewidth{0pt}
\tablecaption{Adopted parameters for all observed Crab glitches.\label{alltbl}}
\tablehead{
	& & \colhead{$t_g$}
	& \colhead{$\Delta\nu_0$}
	& & \colhead{$\Delta\nu_n$} & \colhead{$\tau_n$}
	& \colhead{$\Delta\nu_p$}
	& \colhead{$\Delta\dot{\nu}_p$}
	& \colhead{$\Delta\ddot{\nu}_p$} \\
	\colhead{No.} & \colhead{UT Date} & \colhead{(MJD)} 
	& \colhead{($10^{-7}$ Hz)} 
	& \colhead{$n$} & \colhead{($10^{-7}$ Hz)} & \colhead{(days)}
	& \colhead{($10^{-7}$ Hz)} & \colhead{($10^{-15}$ s$^{-2}$)} 
	& \colhead{($10^{-21}$ s$^{-3}$)} & \colhead{$\tilde{Q}$} }
\startdata
1 & 69-09-30 & 40494   &
        \phn 1.2 $\pm$ .1 &
        1 & \phs\phn 0.7 $\pm$ .1 & 18.7 $\pm$ 1.6 &
        \phn 0.5 $\pm$ .1 &
        \phn $-1.4$ $\pm$ .4 &
	\nodata & 0.58 \\[2ex]
%%\tableline
2 & 75-02-04 & 42447.5 &
        13.2 $\pm$ .2 &
        1 & \phs 10.1 $\pm$ 1.2 & \phd\phn 18 $\pm$ 2 &
        10.2 $\pm$ 1.2 &
        \phn\phd $-92$ $\pm$ 1 &
	\nodata & 0.77 \\
& & & & 2 & $-7.07$ $\pm$ 0.1 & \phd\phn 97 $\pm$ 4 \\[2ex]
%%\tableline
3 & 81-??-?? & $\sim 44900$  &
	\nodata &
	2 & \phn $-2.8$ $\pm$ .1 & \phd 222 $\pm$ 20 &
	\phn 2.8? &
	\phn $-3.8$ $\pm$ .7 &
	\nodata & \nodata \\[2ex]
%%\tableline
4 & 86-08-22 & 46664.4 &
        1.23 $\pm$ .03 & 
        1 & \phs 1.23 $\pm$ .03 & \phn 9.3 $\pm$ .2 &
        \phn 1.1 $\pm$ .1 &
        \phn $-7.1$ $\pm$ 1.6 &
	\nodata & 1.00 \\
& & & & 2 & \phn $-1.1$ $\pm$ .1 & \phd 123 $\pm$ 40 \\[2ex]
%%\tableline
5 & 89-08-29 & 47767.4 &
        $\sim$ 18.5 &
        1 & \phs 22.8 $\pm$ .1 & \phd\phn 18 $\pm$ 2 &
        23.8 $\pm$ .2 &
	\phd $-155$ $\pm$ 2 &
	\nodata & 0.89 \\
& & & & 2 & $-21.1$ $\pm$ .1 & \phd 265 $\pm$ 5 \\
& & & & 3 & \phn\phn\phd $-7$ & \phn 0.8 \\[2ex]
%%\tableline
6 & 92-11-21 & $48947.0 \pm .2$ &
	\phn 3.0 $\pm$ .4 &
	1 & \phs\phn 2.6 $\pm$ .4 & \phn 2.0 $\pm$ .4 &
	\phn 0.4 $\pm$ .1 &
	\phn\phn\phd $-2$ $\pm$ 1 &
	\nodata & 0.87 \\[2ex]
\tableline\\
7 & 95-10-30 & 50020.6 $\pm$ .3 &
	\phn 0.8 $\pm$ .2 & 
	1 & \phs 0.64 $^{+.19}_{-.08}$ & \phn 3.2 $^{+7.3}_{-2.2}$ &
	0.15 $^{+.05}_{-.15}$ &
	\phn $-5.7$ $^{+4.3}_{-2.1}$ &
	\nodata & 0.80 \\[2ex]
%%\tableline
8 & 96-06-25 & 50259.93 $^{+.25}_{-.01}$ &
	$\sim$ 6.6 & 
	1 & \phs\phn 6.6 $\pm$ .1 & 10.3 $\pm$ 1.5 &
	\phn 3.1 $\pm$ .3 &
	\phn\phd $-83$ $\pm$ 6 &
	\phs 0.9 $\pm$ .6 & 0.68 \\
& & & & 3 & \phn $-3.1$ & \phn 0.5 \\[2ex]
%%\tableline
9 & 97-01-11 & 50459.15 $\pm$ .05 &
	\phn 2.3 $\pm$ .1 &
	1 & \phs\phn 2.0 $\pm$ .1 & \phn 3.0 $^{+0.5}_{-1.1}$ &
	0.32 $\pm$ .13 &
	\phn\phd $-18$ $\pm$ 7 &
	\nodata & 0.87 \\[2ex]
%%\tableline
10 & 97-02-10 & 50489.0 $^{+2.5}_{-0.5}$ &
	$\sim$ 0.2 &
	3 & \phn $-0.3$ & \phn 2.2 &
	0.50 $\pm$ .08 &
	\phn $-4.8$ $\pm$ 1.8 &
	$-2.1$ $\pm$ .7 & \nodata \\[2ex]
%%\tableline
11 & 97-12-30 & 50812.9 $^{+0.3}_{-1.5}$ &
	\phn 2.6 $\pm$ .7 &
	1 & \phs\phn 2.4 $\pm$ .6 & \phn 2.9 $\pm$ 1.8 &
	0.17 $\pm$ .05 &
	$-14.2$ $\pm$ .6 &
	\nodata & 0.92 \\[2ex]
%%\tableline
12 & 99-10-01 & 51452.3 $^{+1.2}_{-1.6}$ &
	\phn 2.9 $\pm$ .5 &
	1 & \phs\phn 2.4 $\pm$ .5 & \phn 3.4 $\pm$ .5 &
	\phn 0.4 $\pm$ .1 &
	\phn\phn\phd $-6$ $\pm$ 2 &
	\nodata & 0.83 \\
\enddata
\tablecomments{
Data from the first five glitches are taken from Lyne, Pritchard, \&
Smith 1993. 
}
\end{deluxetable}

\end{document}